\documentclass[aps,prb,twocolumn,floatfix,superscriptaddress]{revtex4-1}
\usepackage{color,graphicx}
\usepackage{amssymb}
\usepackage{graphicx,epstopdf}
\usepackage{hyperref}

\begin{document}

\flushbottom

\title{Observation of a gel of quantum vortices in a superconductor at very low magnetic fields}

\author{Jos\'e Benito Llorens}
\affiliation{Laboratorio de Bajas Temperaturas y Altos Campos Magn\'eticos, Departamento de F\'isica de la Materia Condensada, Instituto Nicol\'as Cabrera and Condensed Matter Physics Center (IFIMAC), Universidad Aut\'onoma de Madrid, E-28049 Madrid,
Spain}

\author{Lior Embon}
\affiliation{Department of Condensed Matter Physics, Weizmann Institute of Science, Rehovot 76100, Israel}

\author{Alexandre Correa}
\affiliation{Instituto de Ciencia de Materiales de Madrid, Consejo Superior de Investigaciones Cient\'{\i}ficas (ICMM-CSIC), Sor Juana In\'es de la Cruz 3, 28049 Madrid, Spain}

\author{Jes\'us David Gonz\'alez}
\affiliation{Facultad de ingenier\'ia, Universidad del Magdalena, Santa Marta, Colombia}
\affiliation{Theory of Functional Materials, Department of Physics, University of Antwerp, Groenenborgerlaan 171, B-2020 Antwerpen, Belgium}

\author{Edwin Herrera}
\affiliation{Laboratorio de Bajas Temperaturas y Altos Campos Magn\'eticos, Departamento de F\'isica de la Materia Condensada, Instituto Nicol\'as Cabrera and Condensed Matter Physics Center (IFIMAC), Universidad Aut\'onoma de Madrid, E-28049 Madrid,
Spain}
\affiliation{Facultad de Ingeniería y Ciencias B\'asicas, Universidad Central, Bogot\'a 110311, Colombia.}

\author{Isabel Guillam\'on}
\affiliation{Laboratorio de Bajas Temperaturas y Altos Campos Magn\'eticos, Departamento de F\'isica de la Materia Condensada, Instituto Nicol\'as Cabrera and Condensed Matter Physics Center (IFIMAC), Universidad Aut\'onoma de Madrid, E-28049 Madrid,
Spain}
\affiliation{Unidad Asociada de Bajas Temperaturas y Altos Campos Magn\'eticos, UAM, CSIC, Cantoblanco, E-28049 Madrid, Spain}

\author{Roberto F. Luccas}
\affiliation{Instituto de Ciencia de Materiales de Madrid, Consejo Superior de Investigaciones Cient\'{\i}ficas (ICMM-CSIC), Sor Juana In\'es de la Cruz 3, 28049 Madrid, Spain}
\affiliation{Instituto de F\'{\i}sica Rosario, CONICET-UNR, Bv. 27 de Febrero 210bis, S2000EZP Rosario, Santa F\'e, Argentina.}

\author{Jon Azpeitia}
\affiliation{Instituto de Ciencia de Materiales de Madrid, Consejo Superior de Investigaciones Cient\'{\i}ficas (ICMM-CSIC), Sor Juana In\'es de la Cruz 3, 28049 Madrid, Spain}

\author{Federico J. Mompe\'an}
\affiliation{Instituto de Ciencia de Materiales de Madrid, Consejo Superior de Investigaciones Cient\'{\i}ficas (ICMM-CSIC), Sor Juana In\'es de la Cruz 3, 28049 Madrid, Spain}
\affiliation{Unidad Asociada de Bajas Temperaturas y Altos Campos Magn\'eticos, UAM, CSIC, Cantoblanco, E-28049 Madrid, Spain}

\author{Mar Garc{\'i}a-Hern{\'a}ndez}
\affiliation{Instituto de Ciencia de Materiales de Madrid, Consejo Superior de Investigaciones Cient\'{\i}ficas (ICMM-CSIC), Sor Juana In\'es de la Cruz 3, 28049 Madrid, Spain}
\affiliation{Unidad Asociada de Bajas Temperaturas y Altos Campos Magn\'eticos, UAM, CSIC, Cantoblanco, E-28049 Madrid, Spain}

\author{Carmen Munuera}
\affiliation{Instituto de Ciencia de Materiales de Madrid, Consejo Superior de Investigaciones Cient\'{\i}ficas (ICMM-CSIC), Sor Juana In\'es de la Cruz 3, 28049 Madrid, Spain}
\affiliation{Unidad Asociada de Bajas Temperaturas y Altos Campos Magn\'eticos, UAM, CSIC, Cantoblanco, E-28049 Madrid, Spain}

\author{Jazm\'in Arag\'on S\'anchez}
\affiliation{Centro At\'omico Bariloche and Instituto Balseiro, CNEA and Universidad de Cuyo, Av. E. Bustillo 9500, R8402AGP, S. C. Bariloche, RN, Argentina}

\author{Yanina Fasano}
\affiliation{Centro At\'omico Bariloche and Instituto Balseiro, CNEA and Universidad de Cuyo, Av. E. Bustillo 9500, R8402AGP, S. C. Bariloche, RN, Argentina}

\author{Milorad V. Milo\v{s}evi\'c}
\affiliation{Theory of Functional Materials, Department of Physics, University of Antwerp, Groenenborgerlaan 171, B-2020 Antwerpen, Belgium}

\author{Hermann Suderow}
\affiliation{Laboratorio de Bajas Temperaturas y Altos Campos Magn\'eticos, Departamento de F\'isica de la Materia Condensada, Instituto Nicol\'as Cabrera and Condensed Matter Physics Center (IFIMAC), Universidad Aut\'onoma de Madrid, E-28049 Madrid,
Spain}
\affiliation{Unidad Asociada de Bajas Temperaturas y Altos Campos Magn\'eticos, UAM, CSIC, Cantoblanco, E-28049 Madrid, Spain}

\author{Yonathan Anahory}
\affiliation{Racah Institute of Physics, The Hebrew University, Jerusalem 91904, Israel}

\date{\today}

\begin{abstract}
A gel consists of a network of particles or molecules formed for example using the sol-gel process, by which a solution transforms into a porous solid. Particles or molecules in a gel are mainly organized on a scaffold that makes up a porous system. Quantized vortices in type II superconductors mostly form spatially homogeneous ordered or amorphous solids. Here we present high-resolution imaging of the vortex lattice displaying dense vortex clusters separated by sparse or entirely vortex-free regions in $\beta$-Bi$_2$Pd superconductor. We find that the intervortex distance diverges upon decreasing the magnetic field and that vortex lattice images follow a multifractal behavior. These properties, characteristic of gels, establish the presence of a novel vortex distribution, distinctly different from the well-studied disordered and glassy phases observed in high-temperature and conventional superconductors. The observed behavior is caused by a scaffold of one-dimensional structural defects with enhanced stress close to the defects. The vortex gel might often occur in type-II superconductors at low magnetic fields. Such vortex distributions should allow to considerably simplify control over vortex positions and manipulation of quantum vortex states.
\end{abstract}

\maketitle

\section*{Introduction}

Quantized vortices in superconductors arrange spatially in structures that bear some similarities with atomic or molecular arrangements. The main difference is that vortices consist simply of single quantized fluxes with repulsive interactions, whereas atoms and molecules have numerous degrees of freedom and allow for bonding. Nevertheless, interactions among vortices can also be varied because these are related to the properties of the superconducting material hosting them\cite{Brandt95,Blatter94}. Thus, solid, liquid or disordered glass, have been observed in vortex matter\cite{PhysRevLett.100.247003,Guillamon2014,PhysRevB.66.020512}. However, a gel with inhomogeneous vortex density distribution caused by a network has not yet been reported. Experiments made in superconductors at low magnetic fields have until now unveiled vortex clustering in hexagonal lattices due to attractive interactions or the intermediate mixed state in single crystals or polycrystalline vortex arrangements and glassy phases in presence of disorder\cite{Babaev05,PhysRevB.66.020512,Prozorov2008,Moshchalkov09,Nishio10,PhysRevB.81.214501,PhysRevB.84.094515,PhysRevB.83.054515,PhysRevLett.105.067003,Brandt2011,Gutierrez12,Reimann2015}. The vortex density remains spatially homogeneus in all these phases. Here we image the vortex configurations at very low magnetic fields in the type-II superconductor $\beta-$Bi$_2$Pd. We find vortex clusters whose distribution has characteristics specific to gels, such as a wide distribution of intervortex separation, covering widely different distances that diverges when decreasing the magnetic field and is characterized by multiple fractal exponents.

The new vortex configuration is dominated by vortices arranging along linear defects, leaving isolated vortices well separated from their neighbors. Such a configuration should be helpful to manipulate topological quantum vortex states, such as those that might arise in the framework of the recently proposed topological superconductivity in $\beta-$Bi$_2$Pd and in iron pnictide superconductors\cite{Lv16,Wang333,Li238,PhysRevLett.120.167001}.

$\beta$-Bi$_2$Pd crystallizes in a tetragonal structure and becomes superconducting below T$_c\approx 5$ K \cite{Imai12,Herrera15}. Zero-temperature critical magnetic fields are H$_{c1}=225$ G and H$_{c2}=6000$ G (yielding coherence length $\xi(0)\approx23.5$ nm and penetration depth $\lambda(0)\approx 132$ nm). The upper critical field anisotropy is small, of order of 10\% \cite{Herrera15,Kacmarcik16,PhysRevLett.120.167001}. Single crystals of this material can be easily cleaved, leaving large stable and atomically flat terraces, where a fully formed s-wave superconducting gap is observed over the entire surface \cite{Herrera15}. Here we undertook a thorough examination of vortex states using SQUID-on-tip (SOT) microscopy\cite{Vasyukov2013,Halbertal2016} and magnetic force microscopy (MFM)\cite{Correa2019} in a large range of magnetic fields, from 1 to 600 G.

\begin{figure*}
	\includegraphics[width=\textwidth]{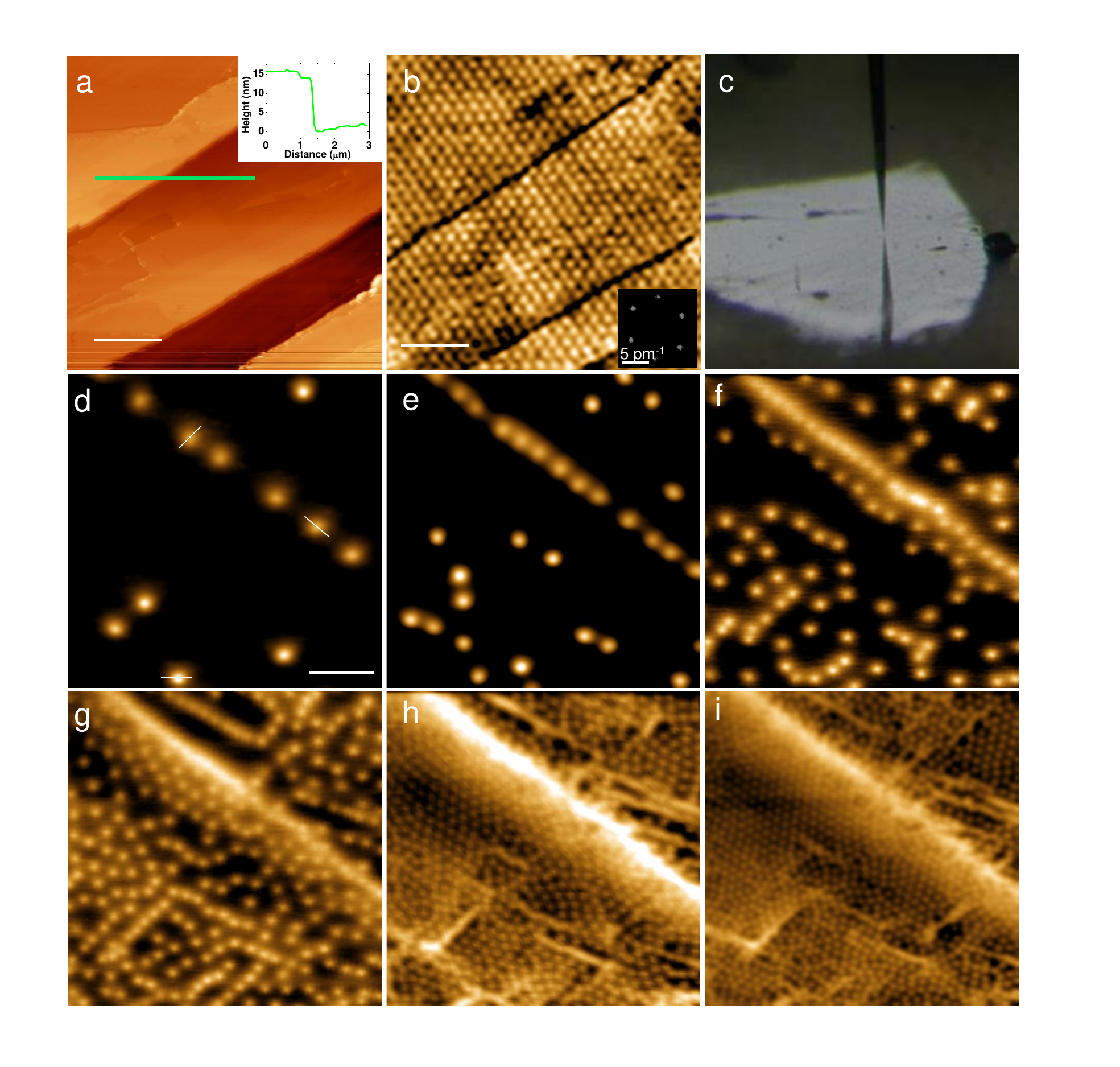}
	\vskip 0cm
	\caption{(\textbf{a}) shows a typical topographic AFM image, with a line profile that show a 15 nm step in the inset (green line). (\textbf{b}) MFM image showing the vortex lattice at 2 K and 300 G and its Fourier transform (inset). The vortex lattice is clearly hexagonal over the whole area. The color code represent the shift of the resonance frequency of the cantilever where white denotes the normal phase and black the superconducting phase. Dark lines and other darker regions in the MFM image are the result of the non-magnetic tip-sample interaction. The scale bar in both images is of 1.4 $\mu$m. (\textbf{c}) Optical image of the SOT at a few tens of $\mu$m from the $\beta$-Bi$_2$Pd surface. The SOT reflection on the surface is visible on the bottom part. (\textbf{d-i}) $20\times 20$ $\mu$m$^2$ SOT images that represent the out-of-plane field $B(x,y)$ obtained after field cooling the sample in magnetic fields of 2 (\textbf{d}), 3 (\textbf{e}), 6 (\textbf{f}), 12.5 (\textbf{g}), 25 (\textbf{h}) and 50 (\textbf{i}) G. The vortex gel is formed below about 20 Oe. The color scale spans 13 G (\textbf{d-f}), 32 G (\textbf{g}) and  27 G (\textbf{h-i}). The scale bar in (\textbf{d}) is for all SOT images and is 4 $\mu$m long. The vortex profiles along the white line in \textbf{d} are shown in Fig. 7}
	\label{figure1}
\end{figure*}

\section{Experimental}

\subsection*{Sample preparation}

$\beta$-Bi$_{2}$Pd single crystals were grown as described in Ref. \onlinecite{Herrera15}. The $\beta$-Bi$_{2}$Pd phase is thermodynamically unstable at room temperature. Crystals need to be quenched from the high temperature $\beta$-Bi$_{2}$Pd phase by immersion in cold water. Structural characterization gives high quality single crystalline samples. The superconducting transition (T$_{c}=5$ K) is sharp in all thermodynamic measurements\cite{Imai12,Kacmarcik16}. However, the residual resistivity, of 18 $\mathrm{\mu\Omega cm}$ is quite high, pointing out that, during the quench, a sizeable amount of defects has been left in the sample. The electronic mean free path is of $\ell\approx$ 15 nm (from residual resistivity \cite{Herrera15}) and the thickness of the sample is of about 0.1 mm.

We prepare the sample surface by cleaving with scotch tape and obtain atomically flat surfaces suitable for scanning probe imaging. We provide a detailed description of defects and of the surface topography in Appendix A. The magnetic field is applied perpendicular to the surface using superconducting coils. The plate like single crystals used here give demagnetizing factors close to one and we usually measure in field-cooled conditions, heating the sample above T$_c$ after each field change.

\subsection*{Magnetic Force Microscopy}
Magnetic force microscopy (MFM) measurements have been performed in a commercial Low-Temperature SPM equipment from Nanomagnetics Instruments, working in the 300 K - 1.8 K temperature range. The microscope is inserted in a superconducting coil\cite{Galvis2015}. Simultaneous atomic force microscopy (AFM) and magnetic MFM images are obtained in dynamic mode, oscillating the cantilever at its resonance frequency, and using the so-called retrace mode for magnetic detection. In this mode, the tip scans twice over the sample surface at two different distances. A first scan is performed at distances of few nm to extract the AFM profile. Then, the tip is retracted by a selected distance ($\sim100$ nm in this work) and the system repeats the same profile obtained in the first scan. Phase shift of the cantilever oscillation during this retrace scan, caused by the long range magnetostatic interaction, is used to build the MFM image. We use commercial MFM tips from Nanosensors (PPP-MFMR) and magnetize these prior to the measurement by applying a magnetic field of 1500 G at 10 K.

\subsection*{Scanning SQUID-on-tip Microscopy}
Scanning SOT images provides high spatial resolution magnetic imaging\cite{Lachman2015,Anahory2016} reaching single-spin sensitivity\cite{Vasyukov2013,Halbertal2016} and enabling detection of sub-nanometer and ultrafast vortex displacements\cite{Embon2015,Embon2017}. The SOT used in this work had an effective diameter of 260 nm, 160 $\mu$A critical current and white flux noise of around 800 n$\Phi_0$ Hz$^{-1/2}$ above a few hundred Hz. The SOT was mounted in a home-built scanning probe microscope with a scanning range of $30\times30$ $\mu$m$^2$ and read out using a series SQUID array amplifier\cite{Huber2001}. The SOT images in this manuscript show an area of $20\times20$ $\mu$m$^2$ with a pixel size of $100\times100$ nm$^2$. The acquisition time was 100 s per image. All measurements were performed at 4.2 K in an open loop mode, at constant height of $\sim$300 nm above the sample.

\subsection*{Ginzburg-Landau numerical simulations}
The numerical simulations were performed within the Ginzburg-Landau (GL) formalism, taking into account the spatially inhomogeneous microscopic parameters. In their stationary form, the dimensionless GL equations for the superconducting order parameter $\Psi$ and the magnetic vector potential ${\bf A}$ read:
\begin{eqnarray}
&& (-i\nabla-{\bf A})^2\Psi = (f({\bf r})-g({\bf r})|\Psi|^2)\Psi, \label{gl1}\\
&& -\kappa(0)^2\nabla\times\nabla\times {\bf A} = \emph{Im}(\Psi^*\nabla\Psi)-|\Psi|^2{\bf A}, \label{gl2}
\end{eqnarray}
where $f({\bf r})=\frac{1-t({\bf r})^2}{1+t({\bf r})^2}$ and $g({\bf r})=\frac{1}{(1+t({\bf r})^2)^2}$ contain the temperature dependence (proposed empirically in Ref. \onlinecite{ginz}, and proven to describe well the standard experimental observations) and the spatially-dependent critical temperature through $t({\bf r})=T/T_c({\bf r})$.
In Eqs. (\ref{gl1}-\ref{gl2}), $\kappa(0)=\lambda(0)/\xi(0)$ is the GL parameter at zero temperature, the distance is measured in units of the coherence length $\xi(0)$, the vector potential $\vec{A}$ in $c\hbar/2e\xi(0)$, and the order parameter $\Psi$ is scaled to its value at zero field and temperature. We apply a finite-difference representation for the order parameter and the vector potential on a uniform 2D (x,y) Cartesian space grid, with spatial resolution of 0.1$\xi(0)$. We use periodic boundary conditions on the simulated rectangular cell $W_{x}\times W_{y}$, in the form ${\bf A}({\bf r}+{\bf b}_{i})={\bf A}({\bf r})+{\nabla}\eta _{i}({\bf r})$, and $\Psi({\bf r}+{\bf b}_{i})=\Psi \exp(2\pi i\eta _{i}({\bf r})/\Phi_{0})$, where ${\bf b}_{i=x,y}$ are the supercell lattice vectors, and $\eta _{i}$ is the gauge potential. Since the sample is exposed to a homogeneous perpendicular magnetic field ${\bf H}=H{\bf e}_{z}$) we employ the Landau gauge ${\bf A}_{ext}=Hx{\bf e}_{y}$ for the external vector potential and $\eta _{x}=HW_{x}y$ while $\eta _{y}=0$. The value of applied magnetic field must always match the integer number of flux quanta (vortices) in the simulated cell, as stipulated by the virial theorem\cite{doria}. In most simulations we deliberately choose $W_x=\sqrt{3}W_y$ that should favor a periodic triangular lattice of vortices, so that every departure from an Abrikosov lattice pattern is easily seen.

\section*{Results}

In Fig.\,\ref{figure1}(\textbf{a}) we show a $7\times7$ $\mu$m$^2$ AFM image taken at 2 K. We observe flat surfaces interspersed with steps of 10 nm height or larger. The vortex lattice shown in Fig.\,\ref{figure1}(\textbf{b}) is measured simultaneously with MFM at 300 G. Fig.\,\ref{figure1}(\textbf{c}) shows an optical image of the SOT near the surface. In Fig.\,\ref{figure1}(\textbf{d-i}) we show image of the local magnetic field $B(x,y)$ acquired using the SOT between 2 and 50 G taken in field-cooled conditions at 4.2 K. At low magnetic fields, Fig.\,\ref{figure1}(\textbf{d,e}), many vortices are located along one dominant line, while, in the rest of the frame, we observe large vortex-free regions. When increasing the magnetic field, vortices cluster along lines and the vortex lattice is formed in between (Fig.\,\ref{figure1}(\textbf{f-i})).

\begin{figure}
\includegraphics[width=\linewidth]{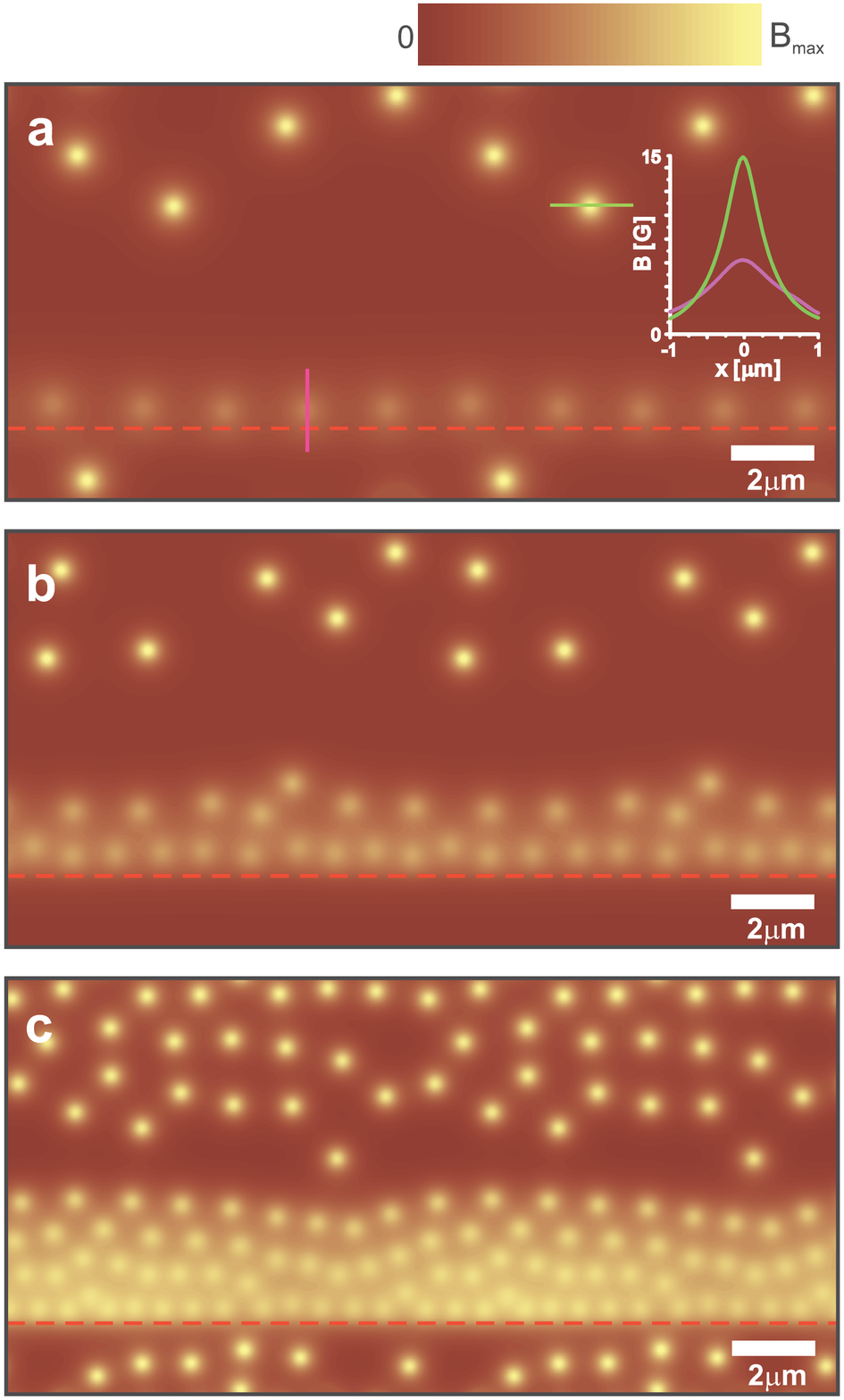}
\caption{Simulation of vortex configurations close to a linear defect. We show vortices as bright spots. To obtain vortex positions, we introduce a critical temperature variation as a function of the position from the defect $\Delta x$ , taking T$_c(\Delta x)$=T$_{c,defect}$+(T$_{c,bulk}$-T$_{c,defect}$)$\frac{(\Delta x)^2}{L_D^2}$, where $L_D$ is the lateral size of the defect. Note that we allow for a slow decay when leaving the linear defect. The position of the defect is marked by a red dashed line. Panels (\textbf{a-c}) show a field of view of $10\times 20$ $\mu$m$^2$, for applied magnetic fields of 2 G, 5 G, and 15 G, respectively. We plot the magnetic induction at a height of 300 nm above the sample surface. This gives a span of $\approx$ 15 G (\textbf{a}), 16 G (\textbf{b}) and 18 G (\textbf{c}) in each image. Inset in (\textbf{a}) shows comparative magnetic profiles of two selected vortices, marked in the main panel by green and magenta lines.}
\label{figGL}
\end{figure}

Note that for vortices located along lines, the value of the magnetic field at the vortex center is smaller than the value we find for vortices located far from lines. This is nicely visible at low fields when vortices are well separated and do not overlap. For example, in Fig.\,\ref{figure1}(\textbf{d}) we see that vortices located along the main linear defect visible in the image (from upper left to middle right) are not as bright as those located elsewhere in the image. This remains in the whole range of magnetic fields (in Fig.\,\ref{figure1}(\textbf{f-i}) there is strong vortex overlap, which produces linear vortex clusters that appear very bright).  When we fit the spatial profile of vortices along lines in the images at low fields we find values for $\lambda$ close to 340 $\pm$ 40 nm, which is about twice the penetration depth found in experiments measuring bulk field penetration with Hall probes\cite{Kacmarcik16}. Further details on the fitting procedure are given in the Appendix B.

We have made numerical simulations of vortex behavior close to linear defects using Ginzburg-Landau theory. We parametrize the simulations according to the values of $\lambda$ measured, that are translated into small changes of T$_c$ close to the defects and weak disorder (i.e. a randomized value of $T_c$). The behavior close to the linear defect is modeled through a parabolic recovery of T$_c$ at a distance of 2 $\mu$m away from the step. This simulates the pinning potential of the linear defect. The obtained vortex configurations over a $10\times 20$ $\mu$m$^2$ area are shown in Fig.\,\ref{figGL}(\textbf{a}-\textbf{c}), for three values of applied magnetic field. This indeed captures the evolution seen in the images. Namely, vortices first occupy locations along the defect, where superconductivity is suppressed, and their magnetic field is weaker than that of vortices found away from the defect. The accumulation of vortices at the defect strengthens repulsion of other vortices in the sample, so that a noticeable vortex-free band is formed between vortices of different brightness in the images. The vortex free band diminishes with increasing magnetic field. Similar vortex free bands are observed in the experiment (Fig.\,\ref{figure1}(\textbf{f-i})).

\begin{figure*} 
	\includegraphics[width=\textwidth]{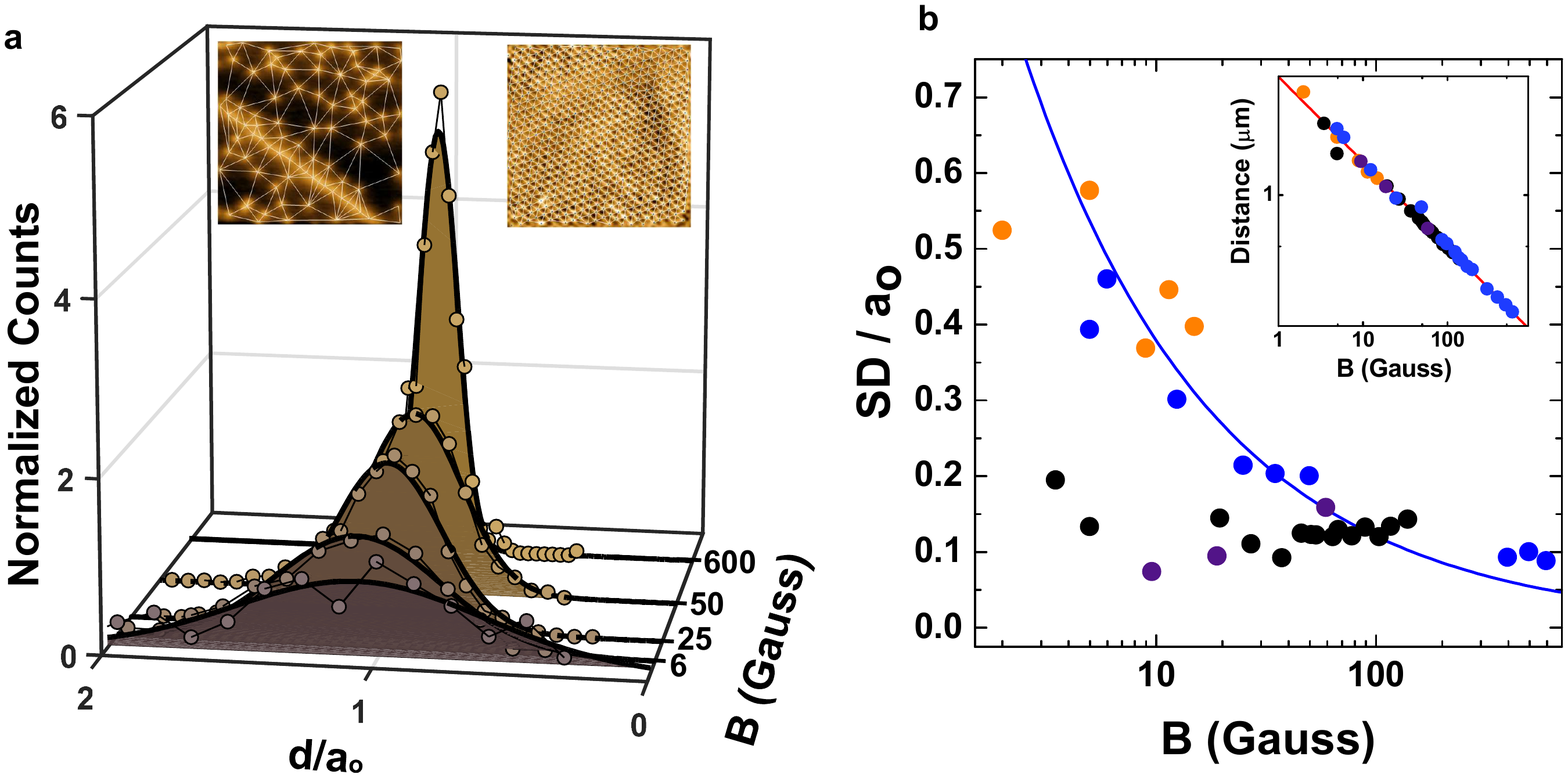}
	\vskip -1.5cm
	\caption{(\textbf{a}) Histograms of the intervortex nearest neighbor distances for different magnetic fields normalized by the integrated area at 6 G, 25 G, 50 G and 600 G. In the insets we show SOT (left inset, image taken at 5 G) and MFM (right inset, image taken at 600 G) images, together with their Delaunay triangulation (white lines). (\textbf{b}) Standard deviation (SD) from histograms in (\textbf{a}), normalized to the intervortex distance $a_0$. Data from SOT and MFM are shown as blue dots. The simulations are shown as orange dots. Black points correspond to the data in the quasicrystalline Bragg glass phase nucleated in pristine cuprate samples and violet points to a disordered lattice in presence of columnar defects. The blue line is 1/$\sqrt{\mu_0H}$. In the inset we show intervortex distance vs the magnetic field for the same images.}
	\label{figure3}
\end{figure*}

The more striking result appears at the smallest magnetic fields. There, vortices are located along lines separated by large vortex free areas (Fig.\,\ref{figure1}(\textbf{d})). We have Delaunay triangulated images and calculated intervortex distances. In Fig.\,\ref{figure3}(\textbf{a}) we show intervortex distance histograms and in Fig.\,\ref{figure3}(\textbf{b}) the standard deviation of intervortex distances. First of all, let us remark that all distance histograms (Fig.\,\ref{figure3}(\textbf{a})) show just a single peak located at $a_0=(\frac{4}{3})^{1/4} \left(\frac{\Phi_0}{B}\right)^{1/2}$ for all magnetic fields. This means that the average intervortex distance follows the Abrikosov rule for a triangular lattice. The magnetic field behavior of the position of this peak is shown in the inset of Fig.\,\ref{figure3}(\textbf{b}) and we can see that its value coincides with $a_0$ also at the weakest magnetic fields. This includes situations where vortices are strongly clustered with large voids in between clusters. At the weakest magnetic fields, the histograms of Fig.\,\ref{figure3}(\textbf{a}) are extremely broad and the distances between vortices become wide-spread. We observe intervortex distances $d$ up to twice $a_0$ and many vortices located much closer to each other than $a_0$. The standard deviation of the histograms as a function of the magnetic field (Fig.\,\ref{figure3}(\textbf{b})) diverges as 1/$\sqrt{\mu_0H}$ below about 80 G.

We compare our results with distance histograms obtained in the same range of magnetic fields in a cuprate superconductor, Bi$_2$Sr$_2$CaCu$_2$O$_8$ in the cases of pristine samples and crystals with a strong and dense distribution of pinning centers (columnar defects)\cite{PhysRevB.93.054505}. In both cases the vortex density is much more homogeneous than in $\beta$-Bi$_2$Pd and the standard deviation remains constant when decreasing the magnetic field (black and violet points in Fig.\,\ref{figure3}(\textbf{b})).

\section*{Discussion}

All our experiments are in field-cooled conditions, so during cooling, vortices are nucleated inside the sample. Our data show that there are locations with vortex pinning determined by linear defects, which provide a scaffold on which vortices arrange and areas in between with just a few vortices. Other mechanisms that can lead to a modified vortex distribution, different than pinning, are discussed in Appendices C and D and can be definitely excluded in $\beta-$Bi$_2$Pd.

Vortices located on the scaffold have a different $\lambda$. We consider stress as a possible mechanism to explain the spatial variation of $\lambda$. Stress builds up close to defects and modifies the superconducting length scales (and the Ginzburg-Landau parameter), producing an effective interaction between the crystal lattice and the vortex lattice\cite{PhysRev.170.470,Johansen00}. Recently, it has been shown that stress induced intervortex interaction can lead to square vortex lattice in tetragonal superconductors\cite{PhysRevB.95.054511}. Such a coupling between crystalline elasticity and superconductivity can be treated using the dependence of T$_c$ with pressure, $dT_c/dP$\cite{PhysRev.170.470,PhysRevB.51.15344,PhysRevB.87.020503,PhysRevB.68.144515}. Generally, with $dT_c/dP>0$, vortices are repelled from places with internal stress and the opposite occurs with $dT_c/dP<0$. The pressure dependence of T$_c$ in $\beta$-Bi$_{2}$Pd was measured in bulk samples\cite{Pristas2018,Zhao2015}, giving $dT_c/dP = -0.025$ K/kbar. T$_{c,bulk}=5$ K for the unstressed region and our SOT measurements are performed at $T=4.2$ K. Thus, a variation in T$_c$ of a fraction of a degree is sufficient to change penetration depth by a factor $\sim$2. If we consider that $\lambda(T) = \lambda(0)/\sqrt{1-(T/T_c)^4}$, we estimate T$_{c,defect}=4.38$ K. According to the study of Ref.~\onlinecite{Pristas2018}, this T$_c$ corresponds to a local pressure of $\sim$ 20 kbar. This subtle local change in T$_c$ is very difficult to detect (we provide further details in the Appendix E).

To analyse further the vortex distributions, we calculate the elastic energy associated to pairs of vortices, $F$, at different locations in our images. We compare the result for vortices located at a step with the elastic energy for pairs of vortices far from the steps. To this end, we use $F=\frac{\phi_{0}^{2}}{4\pi\mu_{0}\lambda^{2}}\log(\kappa)+\frac{\phi_{0}^{2}}{4\pi\mu_{0}\lambda^{2}}K_{0}(d/\lambda)$ for the free energy per unit length of two vortices interacting with each other at a distance $d$ \cite{Tinkhambook}. The first term comes from the energy of superfluid currents, giving the line tension of the vortex, and the second term denotes the interaction energy between vortices. $K_0$ is the order zero modified Bessel function of the second kind. We then calculate $F$ for vortices far from defects using the bulk $\lambda$. For vortices at the steps we use $\lambda$ measured close to defects. This is an approximation since, during a field cooling procedure, the vortex lattice can form closer to $T_c$ where $\lambda$ is significantly larger\cite{Marchevsky97}. However, as we take SOT images at quite high temperature and with an increased $\lambda$, we can still gain useful insight with these approximations. We find that the difference in free energy between the two situations is $\delta F \approx 7 \times 10^{-12}$ Jm$^{-1}$. This result is independent of the intervortex distance, for fields of order of a few tens of G or less. Below $\approx 50$ G the intervortex distances vary from 0.5 to 4 $\mu$m and the second term of the interaction energy remains negligible with respect to the first term. Thus, at low magnetic fields, the intervortex interaction essentially vanishes and vortices behave as nearly isolated entities.

For fields above $\approx 50$ G, the vortex lattice density increases and the previous two-vortex interaction approximation is no longer valid. We take into account the interaction with the first six neighbours arranged in a hexagonal lattice. We can use the same equation as before, but with an interaction term of $6\times\frac{\phi_{0}^{2}}{4\pi\mu_{0}\lambda^{2}}K_{0}(a_0/\lambda)$. The difference in energy between six vortices close to a defect obtained using the two different values of $\lambda$ considered here changes with the intervortex distance. We find that when vortices at the linear defect are closer than about 250 nm, it is no longer energetically favorable to add new vortices in there, so that linear defects saturate and a homogeneous vortex lattice is formed over the whole sample. The field at which we change regime is consistent with the experiment.

\begin{figure}
	\includegraphics[width=\linewidth]{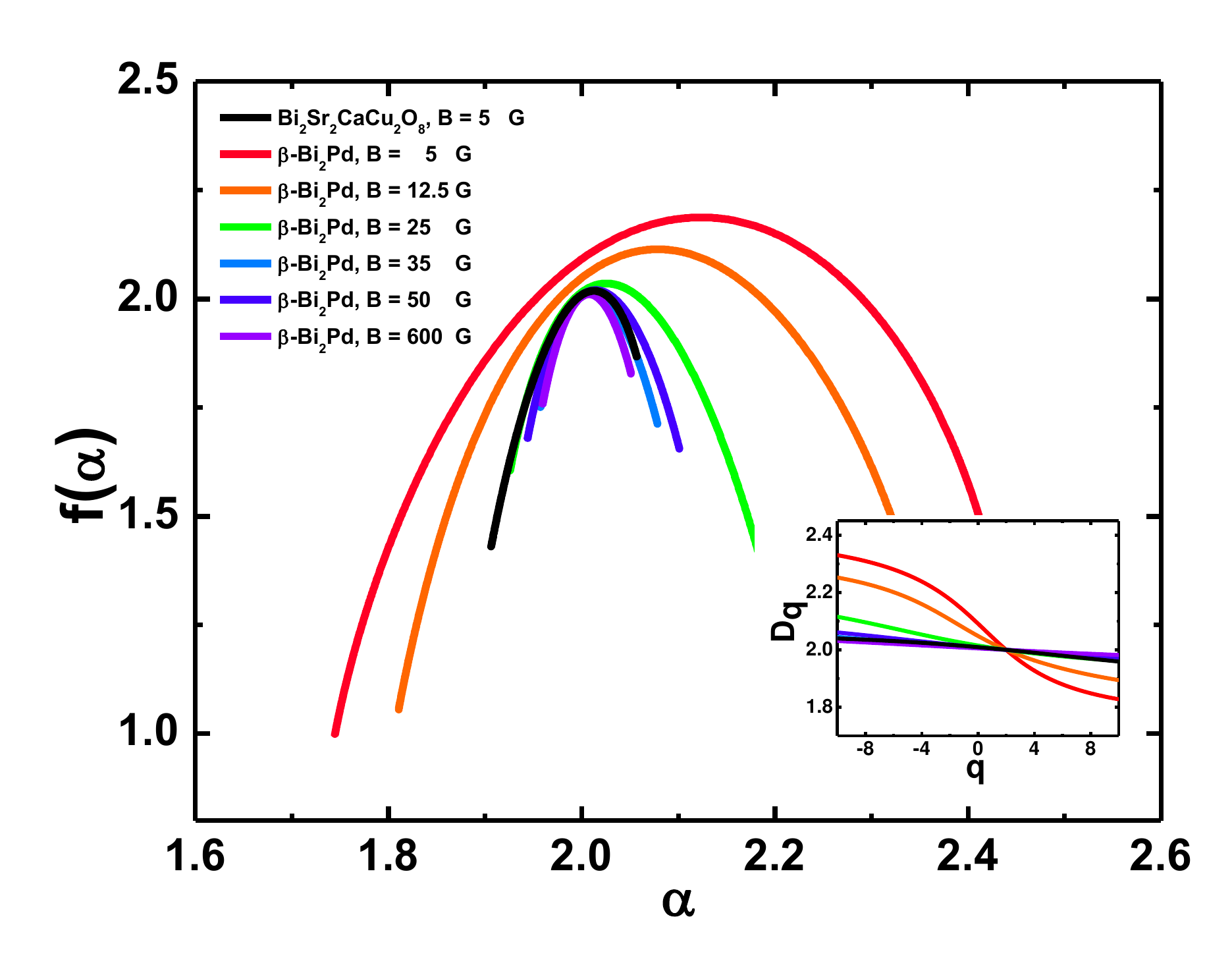}
	\caption{We show in the main panel the distribution of fractal dimensions $f(\alpha)$ in a cuprate superconductor (black line) for 5 G and in $\beta$-Bi$_2$Pd (rest of colored lines) for different magnetic fields from 5 G to 600 G. For the cuprate superconductor, larger images or images at other magnetic fields lead to curves with much smaller dispersion of $\alpha$, centered at $\alpha=2$. In the inset we show the generalized dimension $D_q$ as a function of the set of exponents $q$. Note that the curves strongly vary for low magnetic fields, but remain centered around 2 for high magnetic fields and for the results in the cuprate superconductor. Details on the calculation of multifractal parameters are provided in Appendix F.}
	\label{figure4}
\end{figure}

Thus, our direct imaging of the evolutionary patterns of superconducting vortex configurations in $\beta$-Bi$_{2}$Pd in a broad range of magnetic fields shows that linear pinning centers produce a scaffold that cluster vortices along lines and leave large vortex free areas. At higher fields, when the intervortex distance is much smaller than the separation between defects, a triangular lattice forms within the vortex free areas observed at low magnetic fields.

Another striking insight from our measurements is that the standard deviation of the distribution of vortices diverges for very small vortex densities as the inverse of the square root of the magnetic field. The flux is quantized at each vortex, so that we can assume that $B=\frac{N}{A}\Phi_0$ within the fields of view studied here ($N$ is the total amount of vortices and $A$ the area) and in agreement with the magnetic field dependence of the intervortex distance following the triangular lattice expression (inset in Fig.\,\ref{figure3}(\textbf{b})). In case of a purely one-dimensional row of vortices of length $L$, the average magnetic field along the row goes like $B\propto\frac{N}{L}\Phi_0$ and the intervortex distance diverges as $d_{1{D}}\propto \left(\frac{\Phi_0}{B}\right)$, which gives intervortex distances $d_{1{D}}$ much smaller than $a_0$ if we consider vortex rows in two-dimensional areas. A scaffold of one-dimensional lines distributed within a two-dimensional area $A$ presents a standard deviation of intervortex distances normalized by $a_0$ which increases when decreasing the magnetic field and diverges as 1/$\sqrt{\mu_0H}$, as we observe experimentally.

On the contrary, if we consider a homogeneous random distribution of point pinning centers, the vortices are pinned randomly. As the intervortex distance varies with the vortex density, the deviation of intervortex distances over the average value remains constant when decreasing the magnetic field. This explains the difference between the data in the vortex configurations observed in cuprates and the divergence observed in $\beta-$Bi$_2$Pd in Fig.\,\ref{figure3}(\textbf{b}). Actually, we can understand the behavior in $\beta$-Bi$_{2}$Pd as 1D dominated pinning at very low magnetic fields, which becomes usual 2D or 3D dominated pinning at higher magnetic fields.

In addition, the difference between the vortex distributions observed in $\beta-$Bi$_2$Pd and in the cuprates is that in the former there are strong spatial variations in the vortex density. Thus, there is a parallel with structural glasses and gels that is straightforward. In a similar way that gels are different from glasses because they are created from spatially highly inhomogenoeus amorphous particle arrangements, the vortex gel differs from other vortex configurations, such as the Bose glass or the Bragg glass, by a highly inhomogeneous vortex distribution. There are multiple forms of gels and all share the presence of a scaffold that holds a network of particles. Accordingly, statistical properties of particle distribution and sizes are varied and mimic the properties of the scaffold. Here we have very simple particles, all equal to each other and characterized by carrying exactly a quantized flux, whereas most gels are formed by polymers or other complex structures. The scaffold is also relatively simple, linear defects on a crystalline lattice. Therefore, the statistics of the vortex gel reveals the distribution of these linear defects in the sample.

Often, gels form self-similar or fractal structures. It has been shown that the fractal dimension remains constant with density, unless the density is significantly decreased, in which case the rugosity of constituents increases with decreasing density and the corresponding fractal dimension too\cite{PhysRevB.37.6500}. It is thus interesting to ask whether the distribution of vortices at low magnetic fields also shows similar features. We have calculated the generalized dimension $D_q$ and the multifractal spectrum $f(\alpha)$ (see Appendix F for details on the calculation). The results are shown in Fig.\,\ref{figure4}. Images showing triangular or disordered lattices provide distribution of fractal subsets centered at $\alpha=D_q=2$. When we start observing strong variations in the vortex density, the lacunarity increases, which leads to $f(\alpha)$ that are much broader and whose maximum deviates from $2$. $D_q$ also increases for small values of the multiscaling exponent $q$. Thus, the vortex distributions at small magnetic fields are multifractal, with a probability of fractal subsets that strongly increases when decreasing the magnetic field, leading to a widening of $f(\alpha)$. On the other hand, the maximum of $f(\alpha)$ and the value of $D_q$ for small $q$ is larger than two and increases when the magnetic field is decreased. As we discuss in Appendix A, we find similar fractal properties in images of the linear defects of $\beta-$Bi$_2$Pd. We further show in Appendix G the Voronoi tesselation of vortex images, showing that vortices cluster along the lines and that vortices in between lines are practically isolated.

The outstanding feature of the vortex gel is that the standard deviation of vortex positions diverges for decreasing magnetic fields, suggesting that the magnetic field penetration is neither a Meissner state interspersed with normal areas (the intermediate state \cite{Prozorov2008}), nor an intermediate mixed state with hexagonal vortex clusters (as in pure Nb\cite{Muhlbauer09,Reimann2015}), nor the disordered vortex lattices observed in high T$_c$ cuprate superconductors\cite{Brandt95,Blatter94}. Other arrangements with strong variations in vortex densities are found in conformal crystal arrays of pinning centers and when periodic and random pinning potentials are formed\cite{PhysRevB.79.014504,PhysRevLett.110.267001,Menezes2017}. In the latter case simulations show that the presence of square crystal and hexagonal vortex lattices can lead to one-dimensional fractal structures, in a small range somewhat below the pinning strength range where the combination of random pinning and the crystal lattice is sufficiently strong to form vortex chains\cite{PhysRevB.79.014504}. Our work shows that this actually occurs in a wide range of magnetic fields in presence of one-dimensional pinning centers, leading to vortex distributions with structural properties typical of a gel.

Vortex positions are determined by pinning, and we do not see indications of thermal fluctuations playing a role in the vortex location. Thus, the transition from the normal Abrikosov lattice into the vortex liquid (which can only appear in an extremely small temperature range in $\beta-$Bi$_2$Pd) and the vortex gel should be continuous and be produced by dynamical arrest. Thus, it depends rather on the diffusion process of vortices along the pinning centers than on equilibrium properties of the superconductor. As such, it should appear in all crystalline superconductors with pinning centers having a spatially inhomogeneous distribution at sufficiently low magnetic fields. Vortex lattices showing voids have been indeed observed a few times in some Fe based materials and in the cuprate superconductors\cite{PhysRevB.55.12753,PhysRevB.85.014524,LUCCAS201447}. The inhomogeneous vortex distributions occur either due to twin boundaries in orthorhombic superconductors\cite{PhysRevB.85.014524}, or to highly inhomogeneous pinning centers\cite{LUCCAS201447}. Data were taken at single values of the magnetic field, sometimes two orders of magnitude below the ones we discuss here\cite{PhysRevB.55.12753}. Thus, the divergence of the standard deviation and the multifractal distribution could not be followed with magnetic field nor identified. Nevertheless, the highly inhomogeneous vortex distributions found there suggests that the vortex gel might occur in many superconductors at low magnetic fields.

\section*{Conclusions and outlook}

Our measurements illuminate the interplay of geometric defects and crystalline stress at very low magnetic fields in superconductors, which turns out to be much more varied than previously thought. We reveal novel vortex arrangements at low magnetic fields, governed by organizing principles that combine pinning centers as scaffolds, screening and intervortex interactions.

Screening has been considered to design mechanical resonators that are isolated from the environment for quantum circuits, to produce a scaffold of traps for cold atoms close to the surface of a superconductor or to design magnetic cloaks\cite{PhysRevLett.109.147205,PhysRevLett.111.145304,Gomory1466}. Recent work shows that Majorana modes might be present in superconducting vortices\cite{PhysRevLett.100.096407}. This calls for methods to manipulate vortices and entangle their states to search for non-Abelian statistics\cite{PhysRevLett.114.017001,Wang333}. Superconductors considered are, among others, Fe based materials whose structural properties and superconducting parameters are similar to $\beta-$Bi$_2$Pd and $\beta-$Bi$_2$Pd itself\cite{Dermirdis2011,Dermirdis2013,Wang333,Lv16,Li238,PhysRevLett.120.167001}.

As we discuss in Appendix H, there is mounting evidence for the presence of triplet correlations in $\beta-$Bi$_2$Pd, particularly from experiments that are probing the surface. The question then arises of why we did not observe a half-integer flux quantum in our experiment. The answer might be that triplet correlations could form only close to the surface. Magnetic vortices, however, are threads through the bulk of the material and the surface properties might be masked by the flux quantization from the bulk. It would be very interesting to repeat our experiments in thin films or very thin layers of $\beta-$Bi$_2$Pd. The appearance of half-integer flux quanta is, in view of the recent reports \cite{Sakano15,Iwaya2017,Li238}, quite likely. Vortices might then carry a Majorana fermion, as proposed in Refs. \onlinecite{Caroli64,Hess90,Guillamon08PRB,Alicea2012,Beenakker2013}. In that case, the large intervortex distances found in the range of magnetic fields we study here should be very helpful to facilitate manipulation of vortices and braiding experiments. For example, a couple of vortices located in between lines could be easily moved around each other, as proposed for instance in Refs.\,\cite{PhysRevLett.115.246403,PhysRevLett.122.146803,Xue_2019}.

Vortex manipulation devices remain indeed very difficult to realize in dense vortex lattices at high magnetic fields. It is fundamental to have well separated and isolated vortices to be able to manipulate and entangle vortices (see Appendix B showing isolated vortices in between lines and the evidences for unconventional superconductivity in Appendix H). The vortex gel produces intrinsically areas with flux expulsion and flux concentration.  Our results show that vortices are nearly independent to each other at very low magnetic fields and that their position is locked to the defect structure in the sample. This suggests that vortices in this field range are also highly manipulable, much more than in usual hexagonal or disordered vortex lattices.

\section*{Acknowledgments}
We acknowledge support, discussions and critical reading of the manuscript from Eli Zeldov, who also devised and set-up the SOT system. We also acknowledge critical reading and suggestions of Vladimir Kogan and Alexander Buzdin. Work performed in Spain was supported by the MINECO (FIS2017-84330-R, MAT2017-87134-C2-2-R, RYC-2014-16626 and RYC-2014-15093) and by the Region of Madrid through programs NANOFRONTMAG-CM (S2013/MIT-2850) and MAD2D-CM (S2013/MIT-3007). The SEGAINVEX at UAM is also acknowledged as well as PEOPLE, Graphene Flagship, NMP programs of EU (Grant Agreements FP7-PEOPLE-2013-CIG 618321, 604391 and AMPHIBIAN H2020-NMBP-03-2016 NMP3-SL 2012-310516). Work in Israel was supported by the European Research Council (ERC) under the European Union’s Horizon 2020 research and innovation program (grants No 802952). Y.F. acknowledges the support of grant PICT 2017-2182 from the ANPCyT. E.H. acknowledges support of Departamento Administrativo de Ciencia, Tecnolog\'ia e Innovaci\'on, COLCIENCIAS (Colombia) Programa de estancias Postdoctorales convocatoria 784-2017 and the Cluster de investigación en ciencias y tecnologías convergentes de la Universidad Central (Colombia). I.G. acknlowledges support of European Research Council PNICTEYES grant agreement 679080. M.V.M. acknowledges support from Research Foundation-Flanders (FWO). The international collaboration on this work was fostered by the EU-COST Action CA16218 \textit{Nanoscale Coherent Hybrid Devices for Superconducting Quantum Technologies} (NANOCOHYBRI). 

\section*{Appendix A: Fractographic analysis of cleaved surfaces of {$\beta$-Bi$_2$Pd}}

\begin{figure} [h]
	\centering
		\includegraphics[width=.48\textwidth]{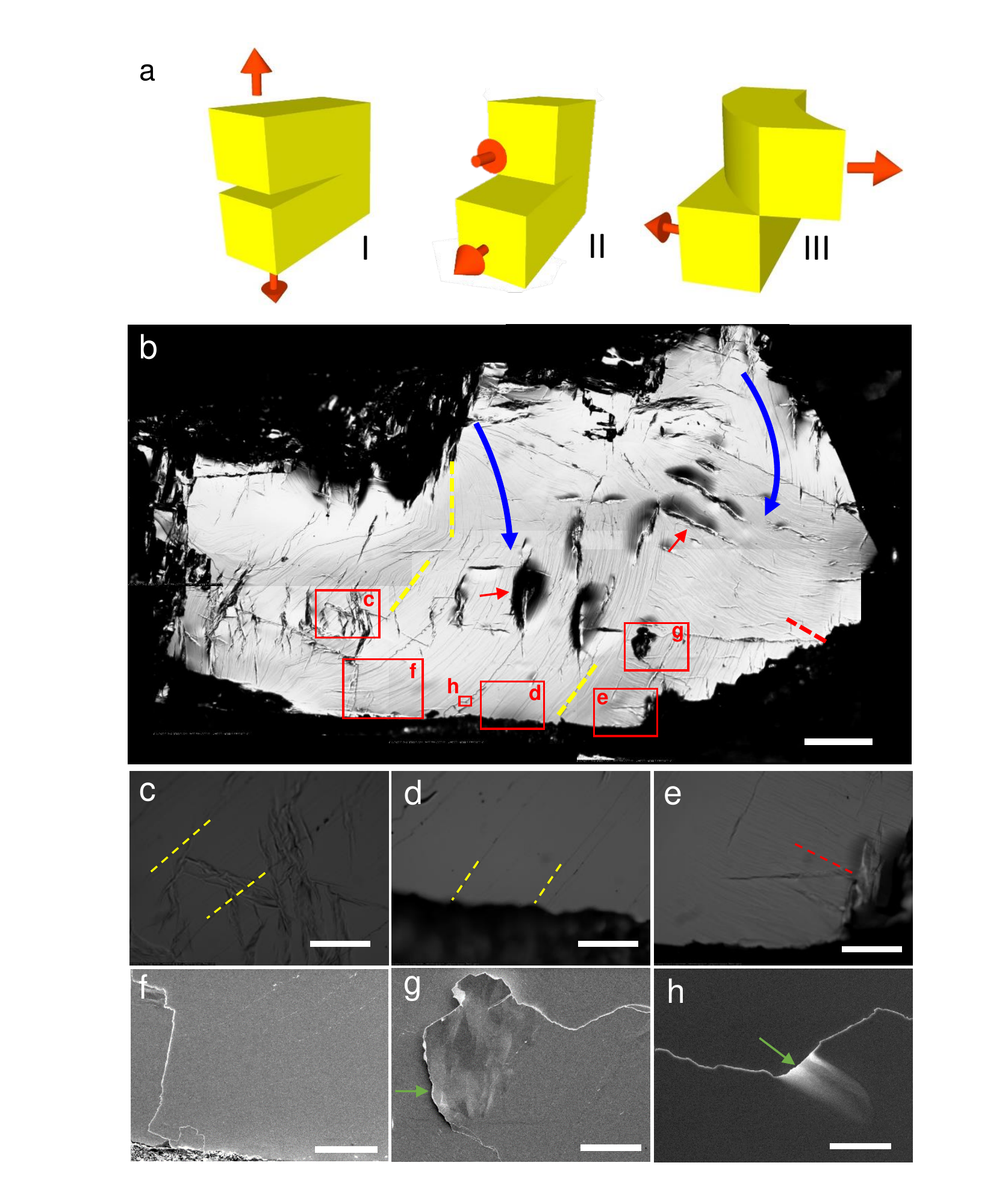}
		\caption{Optical and SEM analysis of fractured surfaces in $\beta$-Bi$_2$Pd. (\textbf{a}) Forces exerted during the cleavage process of a single crystal. In the opening mode (mode I) there is tensile stress normal to the plane of the crack. The sliding mode (mode II) describes shear stress parallel to the plane of the crack and perpendicular to the crack front. The tearing mode (mode III) a shear stress parallel to the plane of the crack and parallel to the crack front. (\textbf{b}) Optical picture of a sample after cleaving. We identify twist hackles (yellow dashed lines), linear features that seem step edges along crystalline directions (red dashed lines) and the debonding path, or the direction where the crack front propagated during fracture (blue lines). (\textbf{c})({\textbf{e}}) Magnified areas marked by red rectangles in (\textbf{b}) taken using an optical camera. (\textbf{f})-({textbf{h}}) SEM images. We mark places where the sample forms fully detached layers by green arrows. Images are at the red rectangles shown in (\textbf{b}) marked by the corresponding letters. Scale bars are of 0.2 mm in (\textbf{b}), 40 $\mu m$ in (\textbf{c}), (\textbf{d}) (\textbf{e}) and (\textbf{g}), 50 $\mu m$ in (\textbf{f}) and 8$\mu m$ in (\textbf{h}).}
		\label{figure9}
\end{figure}

The bonds in the tetragonal structure of $\beta$-Bi$_2$Pd are such that the surface is most likely formed by the square lattice of Bi atoms\cite{Shein13}. The cleaving plane is thus very well defined and lies perpendicular to the c-axis. There are no indications from van der Waals like bonds as in transition metal dichalcogenides\cite{Novoselov10451}---this material has well established three-dimensional electronic properties. Nevertheless, it is a fact that it can be easily cleaved using scotch tape\cite{Herrera15,Kacmarcik16,Sakano15,Iwaya2017}. Cleavage occurs without any residues, as thick sheets of the material are removed when cleaving. The obtained surface is shown in a optical and scanning electron microscope (SEM) images in the Fig.\,\ref{figure9}. The surface is very shiny and has features which are important for the results discussed in the main text.

Cleaving or breaking of a surface occurs through the establishment of a fracture or crack at a few places close to the edges of the sample. The fracture then propagates as a crack front through the whole sample. The action on the sample during fracture consists of tear, shear and compressive forces (called mode I, II and III fracture, respectively, see Fig.\,\ref{figure9}(\textbf{a}) \cite{fracture,MECHOLSKY1995113,PhysRevLett.82.3823}). If the action would occur only along the c-axis, just tear forces that separate layers would be active. However, the competition between elastic energy and surface energy is in-plane anisotropic, leading to crack behavior that depends on the in-plane properties of the material. This occurs irrespective of the anisotropy of the in-plane vs out-of-plane crystalline structure, and is even observed in two-dimensional van der Waals materials\cite{doi:10.1021/acsnano.6b05063}. In three-dimensional crystals this issue is even more important. The sample is of course not perfect and the crack front encounters defects such as small angle grain boundaries\cite{Kermode2013,MARX2010163}. In a cleaving process using scotch tape, shear forces appear easily, because scotch is highly deformable. Shear forces change the direction of crack propagation away from high symmetry crystalline lines. In addition, fracture produces a release of stress that might have been left over during crystal growth (for example by the presence of small temperature gradients). The process of releasing this load is influenced by defects and imperfections in the crystal. All this leads to regions with alternating compressive and tensile stress and causes a twist action (combining shear and tear, modes II and III, see Fig.\,\ref{figure9}(\textbf{a})) that might well propagate far below the surface and influence large areas of the crystal\cite{fracture}.

The features produced by this twist action are called twist hackles and must not necessarily follow a crystalline direction. They rather run parallel to the crack propagation direction. In $\beta$-Bi$_2$Pd we observe features that can be associated to twist hackles. We mark a few of such features by yellow dashed lines in Fig.\,\ref{figure9}(\textbf{b})-(\textbf{d}). Close to the sample edges, twist hackles have a strong tendency to start or arrive to the end of the sample at an angle to the surface, following the crack propagation direction, as we observe in the images (Fig.\,\ref{figure9}(\textbf{d})). The crack propagation direction is also influenced by the direction of the crystalline axis. Thus, there are also a number of linear structures at 45 or so degrees to the twist hackles that can be associated to crystalline axes (we mark a few by red dashed lines in the Fig.\,\ref{figure9}(\textbf{b}), (\textbf{e})). There are furthermore linear features perpendicular to all them. Thus, the images show resulting from twisting efforts produced during fracture. These efforts might produce enough stress to influence locally the superconducting properties.

Let us note that we also identify large wrinkles on the surface (Fig.\,\ref{figure9}(\textbf{b})). The wrinkles appear close to very large defects (broken or open features in the Fig.\,\ref{figure9} (\textbf{a})). A closer analysis using SEM reveals a large number of separated layers close to wrinkles (Fig.\,\ref{figure9}(\textbf{g})). Generally, step edges appear strongly marked in SEM images (Fig.\,\ref{figure9}(\textbf{f}-\textbf{h})), suggesting that some of parts of the sample separate as layers (green arrows in Fig.\,\ref{figure9}(\textbf{g}-\textbf{h})). All this supports the presence of large twisting efforts during fracture.

Our experiments are made close to the center of the sample, in locations showing no large wrinkles and the tip was carefully positioned away from optically visible defects. So that we are far from wrinkles produced during cleaving (red arrows, Fig.\,\ref{figure9}(\textbf{b})). Instead, we perform our experiments close to linear structures due to twist hackles.

\begin{figure} [h]
	\centering
		\includegraphics[width=.48\textwidth]{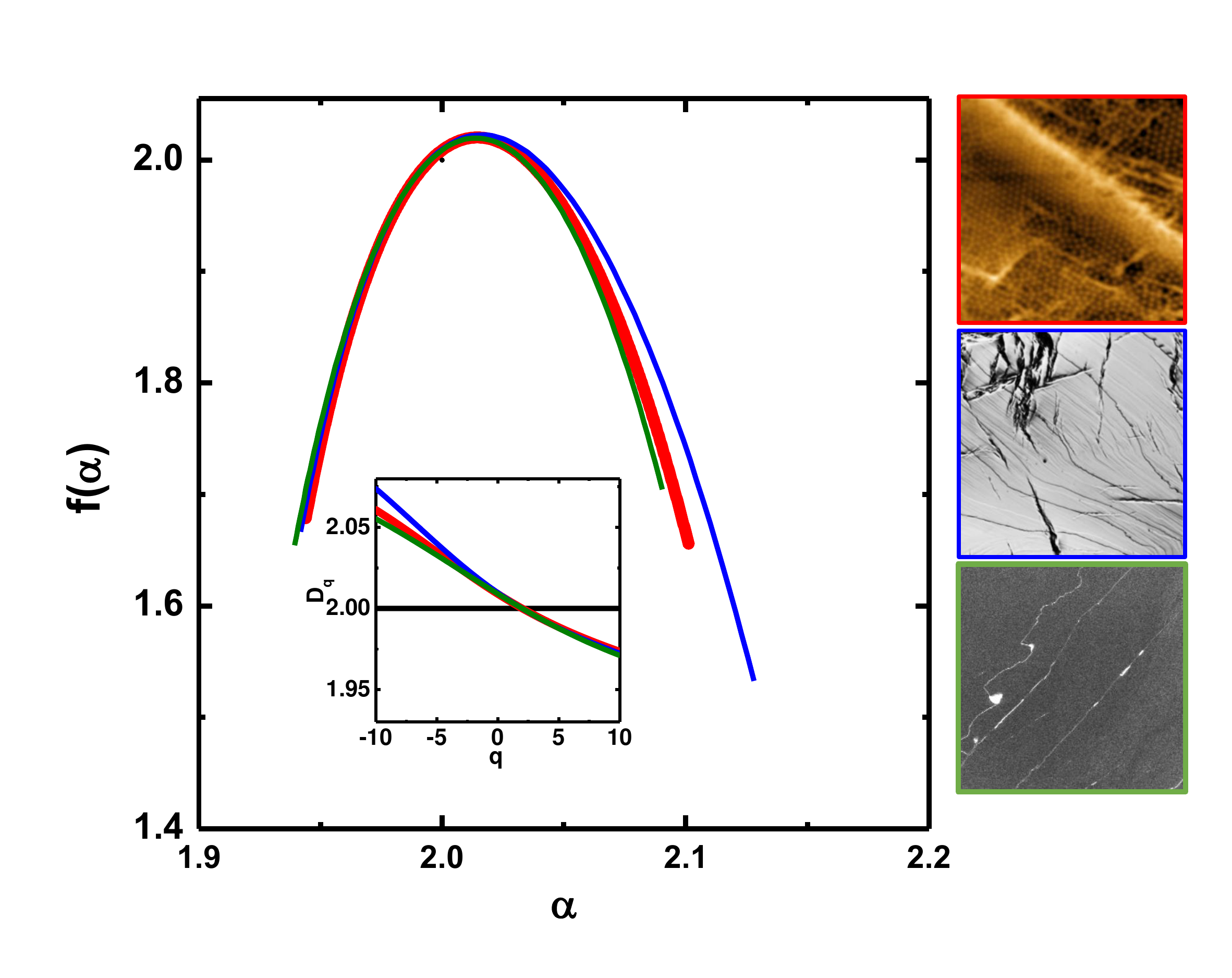}
		\caption{Multifractal properties of vortex lattice and of images of fractured surfaces in $\beta$-Bi$_2$Pd. We show the distribution of fractal dimensions $f(\alpha)$ in the main panel for the vortex lattice image obtained at 50 G (red line), an optical image of the surface (blue line), a scanning electron microscope image of the surface (green line) and an image with random noise (black line). The inset shows the generalized dimension $D_q$ as a function of the set of exponents $q$, also in red (vortex lattice 50 G), blue (optical image), green (scanning electron microscope image) and black (white random noise). Lateral sizes of the images, shown in the panels at the right (with borders of the same colors as in the main panel), are similar, of about 20 $\mu$m.}
		\label{Fig_Topo_Frac}
\end{figure}

It is further interesting to search for similarities in the images showing linear features on the surface and vortex lattice images. To this end, we have taken SEM and optical images of the surface of a similar size as vortex lattice images and calculated their fractal properties (see Appendix F for details). The images of the surface show the linear features that act as a scaffold for vortex pinning, so that we can expect some relation among them. Unfortunately, we cannot analyze exactly the same field of view, because SOT images do not provide surface topographic scans. However, we have found fields of view which are very similar than the ones in vortex lattice imaging, both in terms of density of defects, orientation and distribution. Another caveat is that SEM and optical images provide a continous grey scale corresponding to steps or areas that have a large amount of defects, wheras vortex lattice images are a set of pixelized points (being one at a vortex position and zero elsewhere) that are not joined together. We have thus made a comparison for the image with the largest vortex density, which is the image shown in Fig.\,\ref{figure1}(\textbf{i}) and also increased the contrast of the SEM and optical image by squaring  the grey scale of the image. The result is shown in Fig.\,\ref{Fig_Topo_Frac}. We see that we do obtain the same set of fractal dimensions in vortex lattice images and in the images made using SEM and an optical microscope. Thus, there is a clear relationship between the scaffold of defects and the vortex lattice distribution.

\section*{Appendix B: Fitting procedure of SOT scans}

There are two unknown parameters that are essential to obtain a meaningful fit to a magnetic field profile obtained with SOT.

One is the distance between the sample and the SOT $d_{SOT}$ and the other is the SOT transfer function that converts the measured current that flows in the SOT into a magnetic field ($dI_{SOT}/d{B}$). The latter is normally determined by measuring the response of the SOT to a known applied magnetic field. Here, the presence of the superconducting sample partially screens the applied field. It is therefore somewhat hard to know which part of the applied field is screened by the sample. We thus assume that vortices far from defects, those that appear brightest in the images, have a flux of $\Phi_0$ and $\lambda=$ 186 nm \cite{Kacmarcik16}.

Then, we model the field $B(x,y)$ by a monopole located at $\lambda$ below the surface i.e. at a distance $\lambda + d_{SOT}$ from the SOT. All vortices far from the defects that appear bright in Fig.\,\ref{figure1}(\textbf{d})-(\textbf{e}) are fitted with fitting parameters $d_{SOT}$ and $dI_{SOT}/d{B}$. Averaging results in different vortices we find that $d_{SOT}$= 270 nm. We also obtain the magnetic field scale in the images by obtaining $dI_{SOT}/d{B}$. We considered nine bright and ten less pronounced vortices from Fig.\,\ref{figure1}(\textbf{d})-(\textbf{e}).

\begin{figure} [h]
	\centering
		\includegraphics[width=.45\textwidth]{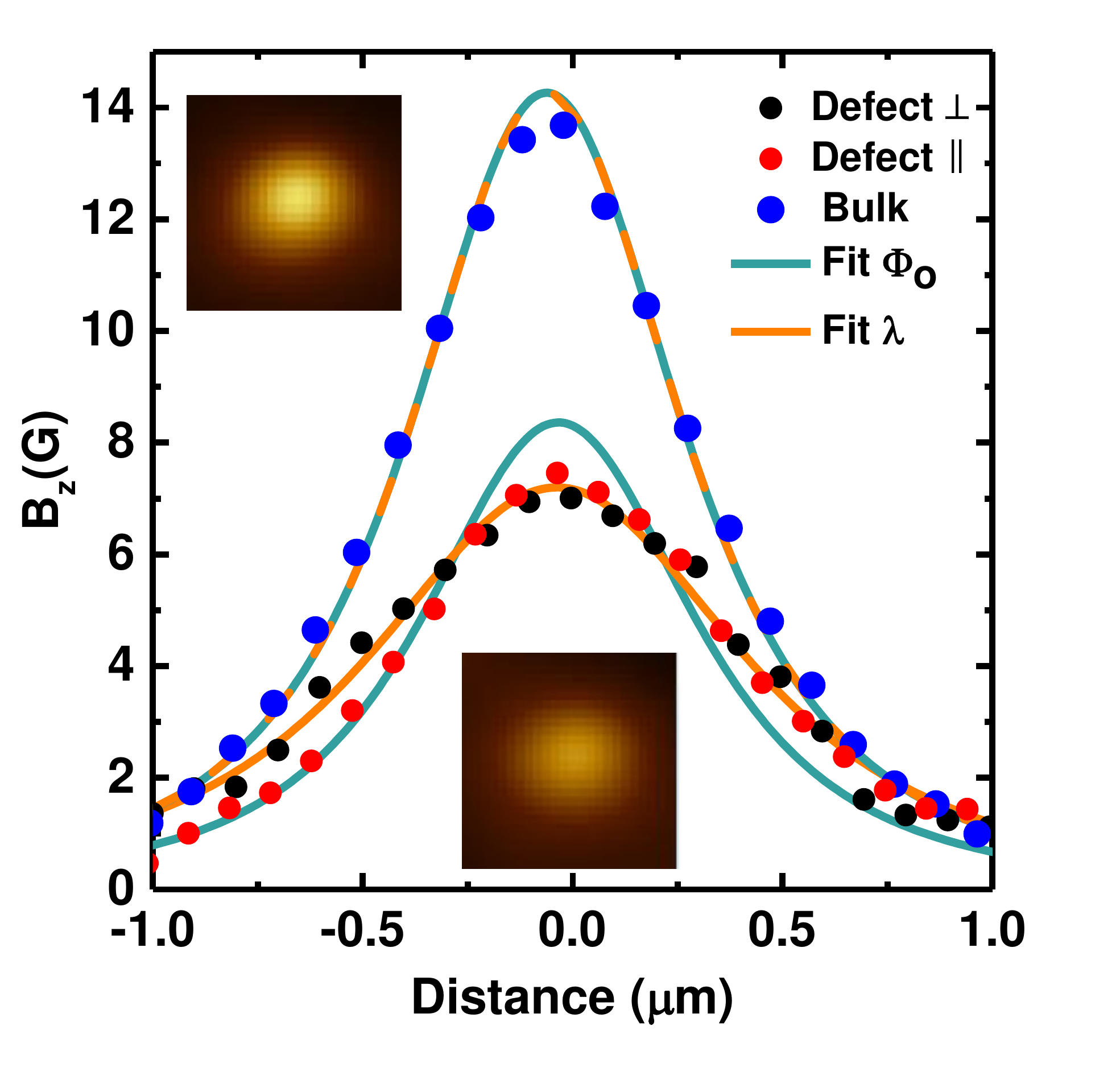}
		\caption{Vortex field profiles $B_z(r)$ along vortices far from a defect (blue points) and close to a defect along the defect (red points) and along the direction perpendicular to the defect (black points). Profiles are taken from Fig.\,\ref{figure1}(\textbf{d}) along the white lines. The fit to the vortex profile leaving the magnetic flux $\phi$ as a free parameter is shown in blue. Vortices far from defects provide an excellent fit with 1.06 $\Phi_0$ where $\Phi_0$ is the flux quantum. However, close to the defects the fit does not provide a good account of the experimental data. By contrast, a fit of the same vortex profiles using the penetration depth $\lambda$ as a free parameter leads to orange lines, which much better account for the experimental data in all situations. We find $\lambda=172$ nm (which is of order of $\sim\lambda$ measured using Hall probes, as discussed in the main article) for vortices far from a defect and $\lambda=360$ nm close to the average value found for vortices at a defect.}
		\label{Fig_Fit2}
\end{figure}

Let us now discuss two representative examples, one showing bright spot and another one a less pronounced spot in the SOT images. The magnetic field profile $B(r)$ across each vortex is shown in the Fig.\,\ref{Fig_Fit2}, together with two different fits. We extract two profiles in orthogonal directions for the vortices on the defect to show that the vortices are isotropic. We obtain similar fits in both direction when vortices are well separated.

We first assumed that the flux is smaller for a vortex located at a defect. To model this situation a fixed $\lambda=186$ nm is considered while the magnetic flux is left as a free parameter. The result is shown in Fig.\,\ref{Fig_Fit2} (green lines). Bright vortices are well described by that value of $\lambda$ and a value for the flux obtained from the fit very close to the flux quantum, at 1.06$\Phi_0$. The magnetic field profile of the vortex on the defect is however not well reproduced.

We then assumed that $\Phi_0$ is the same for all vortices and that lambda differs at different locations. The results are shown in Fig.\,\ref{Fig_Fit2}(black lines). Here, vortices far from defect yield $172$ nm for $\lambda$, which is very close to the reported value\cite{Kacmarcik16}. The vortex close to the defect yields $360\sim\lambda_D$. The values of $\lambda$ stated in the main text are obtained by averaging the results over different vortices. This leads to a constant $\Phi_0$ and is much more consistent with the experiments.

We thus conclude that $\lambda$ is not homogeneous over the sample and is modified at defects in $\beta-$Bi$_2$Pd. 

\section*{Appendix C: Mechanisms producing vortex clustering: Josephson behavior at linear defects, two-gap superconductivity and the intermediate mixed state}

A possible alternative explanation for vortex clustering along lines is that a Josephson-like junction forms step edges with strong pinning and deformation of Abrikosov-Josephson vortices \cite{Gurevich, Roditchev} along such a junction. But the existing low-temperature STM scans of characteristic step edges in $\beta$-Bi$_{2}$Pd do not show evidence of such junctions. It has been also shown that vortex-free regions, called Meissner belts, can appear close to large steps as a result of the Meissner current circulating near steps at the surface\cite{Dermirdis2011,Dermirdis2013}. At the same time, the presence of the steps also imposes boundary conditions on the vortex current which result in an attractive force toward the step edge. The result of these two forces could contribute to some extend to our observations.

We also recall the discussions of possible two-gap superconductivity in $\beta$-Bi$_{2}$Pd \cite{Lv16} and that vortex clustering, stripes and vortex free areas at low magnetic fields have all been observed in the archetypal two-gap superconductor MgB$_{2}$\cite{Moshchalkov09,Nishio10,Gutierrez12}. There, two characteristic distances in the vortex pattern have been reported, with two peaks in the histogram of nearest neighbor vortex distributions. One which is called an intergroup distance and follows $a_{0}$ when varying the magnetic field and an intragroup distance that remains almost constant with respect to the applied magnetic field. It has been argued that the vortex stripes are independent of the crystal lattice and therefore cannot be related to strong pinning. Authors of Refs. \onlinecite{Moshchalkov09,Nishio10,Gutierrez12,Babaev05} relate instead their findings to the magnetic competition of two coexisting gap-condensates in superconducting MgB$_{2}$. The vortex patterns we report here for $\beta$-Bi$_{2}$Pd are different. At very low fields, the patterns do contain vortex stripes, clusters and vortex free regions. However, the intervortex distances do not cluster around two values as seen in MgB$_2$. Instead, vortices arrange in lines along crystalline defects. As convincingly shown in Refs. \onlinecite{Herrera15,Kacmarcik16,Zheng17}, $\beta$-Bi$_{2}$Pd is probably a single-gap superconductor, so vortex clustering cannot be associated to the hybridization of multiple gaps.

Vortex clustering has been seen in single-gap superconductors as well, with $\kappa\gtrsim\sqrt{\frac{1}{2}}$. The intermediate mixed state consists of clusters of vortices with widely differing intervortex distances that are often smaller than $d_{Abrikosov}$. Vortex lattices at fields close to or below H$_{c1}$ imaged using magnetic decoration, scanning Hall-probe microscopy, and/or studied using small-angle neutron scattering often show a mixed intermediate state characterized by strong magnetic field gradients inside the sample, due to flux-free areas coexisting with areas having vortices inside\cite{Trauble68,Hubener01,Fasano08,Brandt87,Aston71,Brandt2011,zrb12}. In high quality single crystals, such as e.g. Nb with $\kappa=1.1\sqrt{\frac{1}{2}}$, flux-free regions coexist with domains of vortex lattice \cite{Brandt2011}, where the shape of domains resembles the intermediate state found in type I superconductors\cite{Prozorov2008}. Small-angle neutron scattering finds the intervortex distance inside those domains exactly as expected at H$_{c1}$\cite{Muhlbauer09,Reimann2015}. In our experiments, the magnetic flux integrated over large areas provides a magnetic induction of the same order in presence of an ordered vortex lattice and in presence of strong vortex clustering. Vortices are nucleated in the sample during the field-cooled procedure, and remain pinned even at very low magnetic fields, in spite of the strong field gradients produced when cooling below H$_{c1}(T)$. Such a feature has never been previously reported, to our knowledge, at low magnetic fields and in presence of strongly inhomogeneous vortex distributions. 

\section*{Appendix D: Fluctuations vs pinning}

The elastic displacement of vortices $u(0)$ caused by a pinning force $F$ depends on the magnetic field and temperature following $u(0)\approx F(\frac{4\pi}{B\Phi_0})^{1/2} \mu_0 \frac{\lambda^2}{(1-\frac{B}{B_{c2}})^{3/2}}\approx F(\frac{4\pi}{B\Phi_0})^{1/2} \mu_0 \frac{\Phi_0 ln\kappa}{B_{c1}(1-\frac{B}{B_{c2}})^{3/2}}$, for an isotropic superconductor with penetration depth $\lambda$ ($\mu_0$ is the magnetic permeability)\cite{Brandt87,Blatter94,doi:10.1002/pssb.2220790210}. Close to steps, the increased $\lambda$ thus yields an increased displacement of vortices $u(0)$. In addition, $B_{c1}\approx \frac{\Phi_0}{4\pi\lambda^2}(ln\kappa+0.5)$ is decreased, whereas $B_{c2}=\frac{\Phi_0}{\pi \xi^2}$ remains similar, provided that $\xi$ is limited by the mean free path in the dirty limit regime.

Vortices suffer positional changes due to thermal activation, which can be discussed using elastic theory of the vortex lattice\cite{Brandt87,Blatter94,RevModPhys.90.015009}. The changes in the position are related to the temperature following $<u^2>\approx d_0^2 (\frac{3G_i B}{\pi^2B_{c2}(0)})^{1/2}\frac{T}{T_c}\frac{\lambda^2(T)}{\lambda^2(0)}((1-B/B_{c2})^3ln(2+\frac{1}{\sqrt{2B/B_{c2}}}))^{-1/2}$ ($a_0$ is the Abrikosov intervortex distance and $G_i$ the Levanyuk-Ginzburg number)\cite{Brandt87,Blatter94,PhysRevB.42.906,PhysRevB.41.8986,PhysRevB.39.136,PhysRevLett.63.1106}. The fluctuation range is very small, because in any event $G_i$ likely small in $\beta-$Bi$_2$Pd. Anisotropy enhances fluctuations but there is practically no anisotropy in $\beta-Bi_2Pd$\cite{Herrera15,Kacmarcik16}. However, fluctuations are enhanced for large wavelengths, by taking into account nonlocal elastic interactions, and in presence of pinning induced disorder due to the mobility of dislocations and defects of the flux lattice\cite{PhysRevB.41.1910,PhysRevB.42.9938,PhysRevB.48.6539,PhysRevB.34.6514,RevModPhys.90.015009}. Vortex nucleation occurs here very close to the zero field $T_c$ for very low magnetic fields, which would suggest that fluctuations might play a role.

Vortex entry occurs through thermal diffusion for long wavelengths $k$\cite{Brandt87,Blatter94}. There are two processes of vortex diffusion, giving an exponential decay $exp(-\Gamma_1 t)$ with rates $\Gamma_1$ and $\Gamma_2$\cite{Brandt87,PhysRevLett.71.3541}. First, through the lattice compression, giving $exp(-\Gamma_1 t)$ with $\Gamma_1 \approx \frac{B}{\mu_0B_{c2}}\rho_Nk^2$ with $\rho_N$ being the normal state resistivity. For very low magnetic fields of a few Gauss, and taking $\rho_N\approx 10^{-6}\Omega m$, we find  $\Gamma_1\approx 3 10^2 L^2$ in s, with $L$ the wavelength in meters. Even for relatively long wavelengths of order of the sample size, this process is fast, in the $\mu s$ range or below. Second, through shear stress, with a rate $\Gamma_2\approx c_{66} k^2$. This process is much slower, because the shear modulus $c_{66}$ is small. If we take $c_{66}\approx \frac{B\Phi_0}{16\pi \lambda^2\mu_0}$ we can find $c_{66}\propto \frac{1}{L^2}$ and time scales three order of magnitude above estimation, but still of the order of a small fraction of a s. A typical cooling procedure after applying the magnetic field is certainly much slower, of the order of one s.

Therefore, we conclude that we do not identify a strong effect of fluctuations and that we are instead observing the sole effect of pinning in our experiments.

\section*{Appendix E: Temperature dependence of the MFM images}

\begin{figure} [h]
\centering
\includegraphics[width=0.48\textwidth]{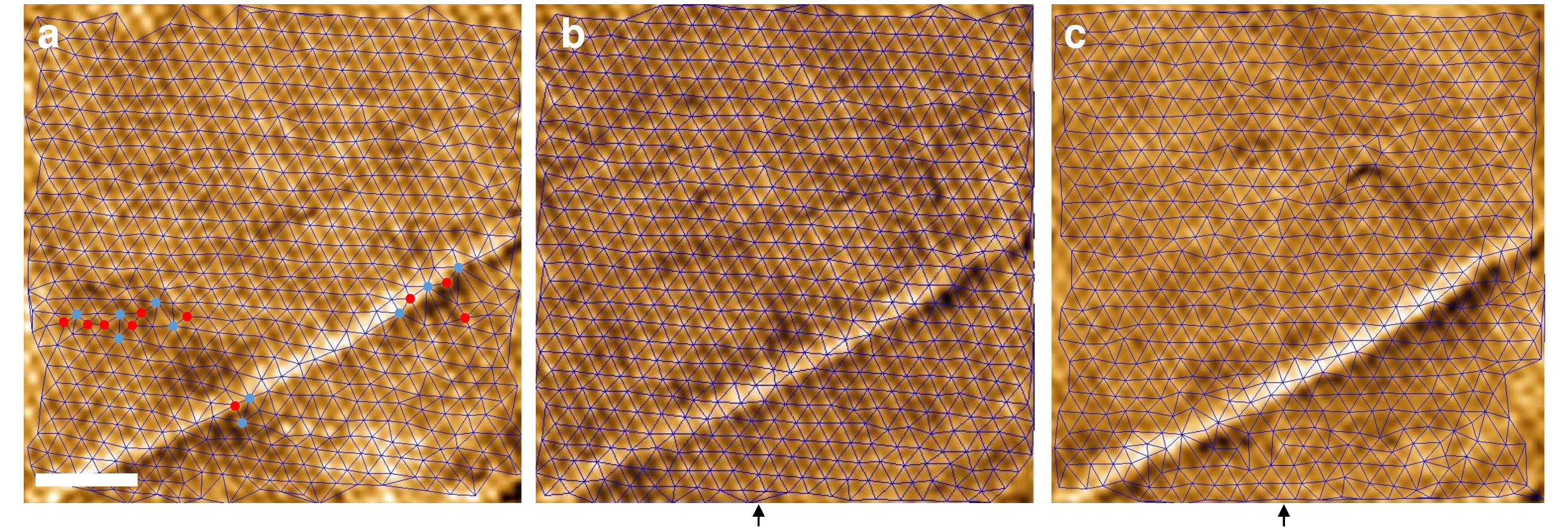}
\caption{ {\bf Behavior of the hexagonal vortex lattice as a function of temperature measured with MFM.} (\textbf{a})-(\textbf{c}) images taken at 2.75 K, 3.75 K and 4.5 K, respectively at 300 G. The color scale represents the observed frequency shift. Scale bar is 1 $\mu m$. Blue lines are the Delaunay triangulation of vortex positions. Blue and red points in (\textbf{a}) highlight vortices with seven and five nearest neighbors respectively. The dark arrow at the bottom highlights the position of the vertical line discussed in the text.}
\label{figure5}
\end{figure}

The evolution of the flux line lattice with increasing temperature is shown in Fig.\,\ref{figure5}. In the MFM experiment, we observe the hexagonal vortex lattice up to temperatures of 4.5 K at a magnetic fields of 300 G. We mostly observe defect free vortex lattices and see no particular increase in the defect structure of the vortex lattice when increasing temperature. In Fig.\,\ref{figure5}(\textbf{a}) there are a few defects in the vortex lattice (vortices showing seven or five nearest neighbors). These are washed out when increasing temperature (Fig.\,\ref{figure5}(\textbf{b}),(\textbf{c})). Thus, the range of vortex liquid is very small in $\beta-$Bi$_2$Pd.

The diagonal line located at the bottom of the image and visible at all temperatures is probably caused by a crosstalk between charging effects between tip and sample and the magnetic signal close to a large step. The crosstalk between the two signals should not be temperature dependent and therefore we expect variations in this feature to be related to temperature induced variations in the magnetic properties. Part of this feature might thus be associated to a decrease in $T_c$ along this line. The crosstalk makes it however very difficult to make any quantitative estimation of $T_c$. On the other hand, the image taken at the highest temperatures (Fig.\,\ref{figure5}(\textbf{b}),(\textbf{c})) show a whitish vertical line highlighted by a black arrow. That line is absent in other images as shown and might result from a change in T$_c$ along this line. With our present resolution in MFM, SOT and available STM experiments\cite{Herrera15,Kacmarcik16} we cannot be more quantitative. SOT suffers from problems with temperature control and the estimated change in T$_c$ corresponds to a change in the gap size of ~50 $\mathrm{\mu V}$ which is impossible to detect in STM measurements taking into account the small but finite gap distribution close to 50 $\mathrm{\mu V}$ found in this compound\cite{Herrera15,Iwaya2017}.

\section*{Appendix F: Details on the calculation of the fractal properties}

To characterize a multifractal system, we mainly use two functions $f(\alpha)$ vs $\alpha$ and $D_q$ vs q.

$f(\alpha)$ vs $\alpha$ is the multifractal singularity spectrum and is typically a concave downward curve with a maximum and a certain width. The $\alpha$ value corresponding to the maximum in $f(\alpha)$ gives the fractal dimension. For example, if we consider pixels consisting of zeroes and ones distributed over a square, we can find different results, depending on the distribution of these pixels. A random distribution of pixels provides a small concave downward curve which is only defined very close to $\alpha=2$ and where $f(\alpha)$ is nearly constant. A monofractal distribution of pixels gives a maximum of $f(\alpha)$ at the fractal dimension. A multifractal distribution of pixels gives a concave downward $f(\alpha)$ defined over a wide range of $\alpha$.

$D_q$ vs q gives the generalized dimension for the set of scaling exponents q. The scaling exponents are used to highlight different regions of the image with more or less concentration of pixels. In random and in monofractal images, $D_q$ is a flat line because the dimension does not depend on q. In a multifractal, the dimension changes with the area, and thus with the scaling exponent q. We obtain a sigmoidal curve for $D_q$ vs q. q can be varied in the range of $ [\infty, -\infty]$ but for the implementation in the calculation, the limits depend on the convergence of the curve $D_q$ vs q for different ranges of q. In our case a convergence in $D_q$  was obtained in the range of $[10, -10]$.

\begin{figure} [h]
\centering
\includegraphics[width=0.49\textwidth]{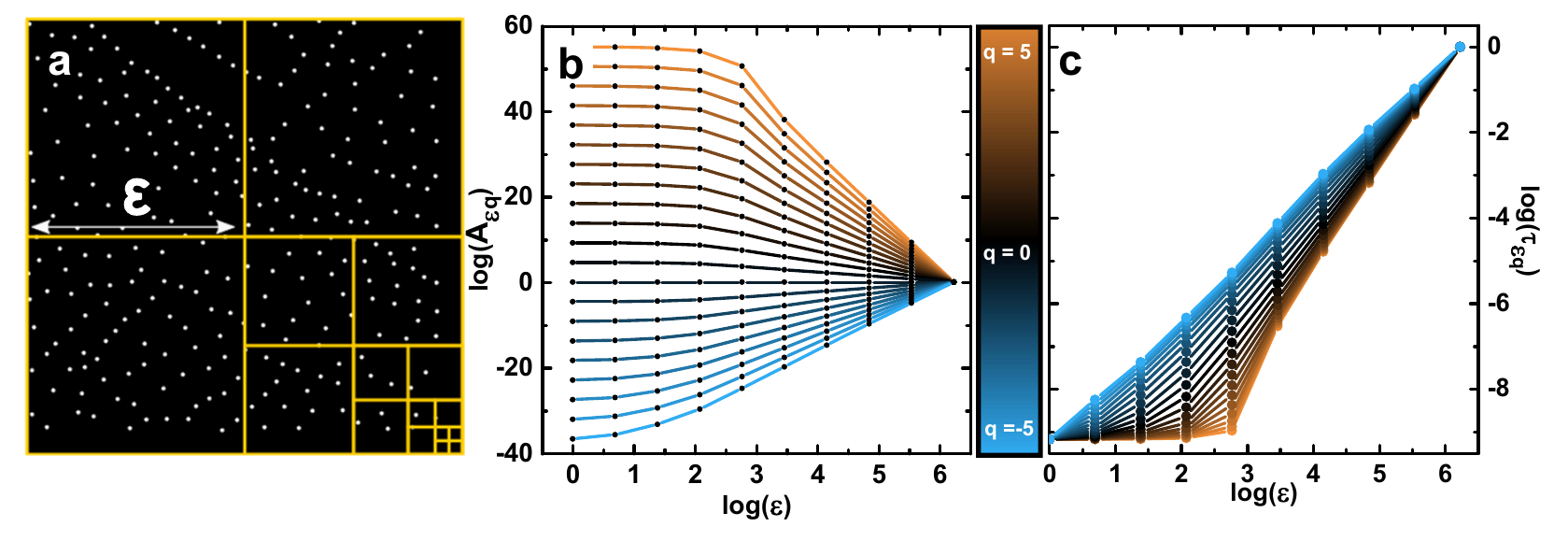}
\caption{ In (\textbf{a}) an example with vortex positions indicating how boxes decrease its size. In (\textbf{b}) we show $log(A_{\epsilon q})$ vs $log(\epsilon)$ curves in a set from q=-5 to q = 5. The curves have a clear slope for large $log(\epsilon)$ and become flat for low $log(\epsilon)$. The change of slope is due to the density of points. We only take in account the points before the change of slope. This gives us $\alpha_q$. In (\textbf{c}) we show $log(\tau_{\epsilon q})$ vs $log(\epsilon)$ curves at the same set of q as in (\textbf{b}). The same behaviour with the size arises and we treat it similarly. The slope of this curve gives us $\tau_q$.}
\label{FigFracCalc}
\end{figure}

To go into the details of how to obtain $f(\alpha)$ and $D_q$, we first remind a few aspects of the calculation we have made. We used the box counting method \cite{Chhabra1989}. For a given binary matrix of points, as \autoref{FigFracCalc}(\textbf{a}), we calculate the number of points, $m_i({\epsilon})$, in each box  of length $\epsilon$, and compute the probability of finding a white pixel in each box with:

\begin{equation}
  P_i(\epsilon) = \frac{ m_i(\epsilon)}{\sum_{i}^{N_{i(\epsilon)}}m_i(\epsilon) }
\end{equation}

being  $N_{i(\epsilon)}$ the number of boxes with length $\epsilon$ containing at least one point. Now we introduce the set of exponents q, which provide the dimensions in the multifractal spectra. We calculate:

\begin{equation}
I_{q\epsilon} = \sum_{i}^{N_{i(\epsilon)}}P_i(\epsilon)^q
\end{equation}

\begin{equation}
\mu_{qi(\epsilon)} = \frac{P_i(\epsilon)^q}{I_{q\epsilon}}
\end{equation}
where $I_{q\epsilon}$ represents how the pixels are distributed in space. The smaller $I_{q\epsilon}$, the larger the homogeneity in the number of pixels inside the boxes of same $\epsilon$.  $\mu_{qi(\epsilon)}$ is equivalent to  $P_i(\epsilon)$ but taking into account the different behaviour of scaling exponents q. Then we calculate
\begin{equation}
A_{\epsilon q} = \sum_{i}^{N_{i(\epsilon)} }\mu_{qi(\epsilon)}P_{i(\epsilon)}
\end{equation}     
    which gives the $\alpha_q$ value by calculating the slope of $log(A_{\epsilon q})$ vs $log(\epsilon)$ as shown in Fig.\,\ref{FigFracCalc}(\textbf{b}). When a discrete set of points is given, as in the case of vortex image, only the box with a size of $\epsilon$ larger than minimum distance between first neighbours is taken into account. That is why the curves of Fig.\ref{FigFracCalc}(\textbf{b}-\textbf{c}) flatten at small $\epsilon$. We also compute
    \begin{equation}
    \tau_{q\epsilon} = \frac{\sum_{i}^{N_{i(\epsilon)}}P_i(\epsilon)^{q-1}}{N_\epsilon}
    \end{equation}
shown in Fig.\,\ref{FigFracCalc}(\textbf{c}). From the slope of $log(\tau_{\epsilon q})$ vs $log(\epsilon)$ we obtain $\tau_q$. Finally we can calculate 
   
\begin{equation}
f(\alpha) = \alpha_qq-\tau_q
\end{equation}   

\begin{equation}
 D_q = \frac{\tau_q}{q-1} 
\end{equation}

To calculate D$_q$ and $f(\alpha)$ for the vortex lattice images (Fig.\,\ref{figure1} ), we have first searched all vortex positions and calculated the images with one at a vortex and zero elsewhere. This leads to maps of pixels with values zero and one. The results are shown in Fig.\,\ref{figure4}. Images showing triangular or disordered lattices provide a distribution of fractal subsets centred at $\alpha = D_q = 2$. When we start observing variations in the vortex density, multifractality increases, which leads to $f(\alpha)$ that is much broader and whose maximum deviates from 2. D$_q$ also increases for small values of the multi-scaling exponent q. Thus, the vortex distributions at small magnetic fields are multifractal, with a probability of fractal subsets that strongly increases when decreasing the magnetic field, leading to a widening of $f(\alpha)$ and a considerable dependence of D$_q$ on q. 

\color[rgb]{0,0,0}

\section*{Appendix G: Voronoi tesselation of the images}

\begin{figure} [h]
\centering
\includegraphics[width=0.48\textwidth]{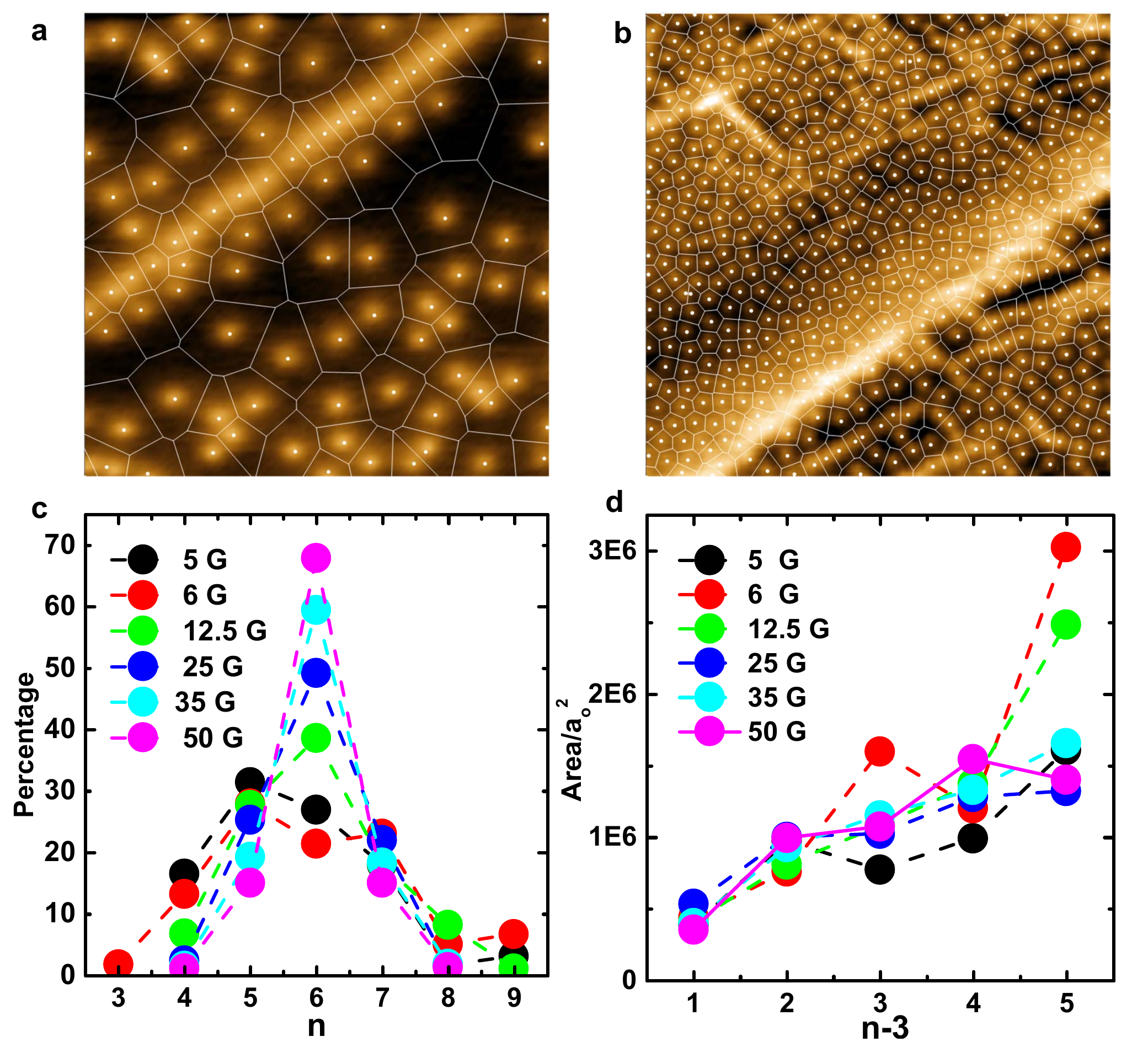}
\caption{Voronoi tesselation of the vortex lattice. (\textbf{a})-(\textbf{b}) vortex lattice images and the Voronoi tesselation (white lines). The position of each vortex is shown by a point. (\textbf{c}) percentage of cells with the number of sides $n$ for different magnetic fields (shown in the legent). Dashed lines are a guide to the eye. (\textbf{d}) Average over all areas occupied by cells having sides $n$ for images at different magnetic fields (shown in the legend).}
\label{FigVoronoi}
\end{figure}

Using the Delaunay triangulation, we have also calculated the Voronoi pattern of the vortex lattice images. Each vortex is then located inside a cell. We calculate the number of sides of each Voronoi cell (in the hexagonal lattice, each vortex cell has 6 sides) and the area, compared to the result in a perfect hexagonal lattice ($a_0^2$). The Voronoi patterns and the evolution of number of sides and area is shown in Fig.\,\ref{FigVoronoi}. At small magnetic fields, the number of sides of the cells significantly deviates from 6 (Fig.\,\ref{FigVoronoi}(\textbf{a}),(\textbf{c}). More cells have less than 6 sides, although the distribution flattens out considerably. Cells with less than 6 sides corresponds to vortices aligned along a row, where vortices are located in cells with mostly 4 sides. On the other hand, vortices in between rows have six sides, or more when they do not have many neighbors. As can be expected (von Neumann's law), the area occupied by vortex cells increases with the number of sides (Fig.\,\ref{FigVoronoi}(\textbf{d}) for all magnetic fields). However, for low magnetic fields, the cells with more than 6 sides also have a larger area. This shows that vortices tend to be isolated in between rows and lines.

\section*{Appendix H: Topologically non-trivial properties of the bandstructure}

Measurements of angular resolved photoemission in the normal phase above T$_c$ suggest non-trivial topological behavior at the surface of $\beta$-Bi$_{2}$Pd \cite{Sakano15,Iwaya2017}. The band-structure observed by photoemission coincides with calculations and has mixed contributions from Pd $4d$ and Bi $6p$ orbitals that give three main sheets of convolved geometries that partially overlap with each other\cite{Shein13,Coldea16}. Photoemission reveals a Dirac cone well below the Fermi level\cite{Sakano15}. Spin resolved measurements provide polarized bands close to the Dirac cone. The same authors suggest that topologically non-trivial spin polarized bands crossing the Fermi level might rise up to the surface. The STM experiment of Ref. \onlinecite{Iwaya2017} provides indications for a small triplet component appearing at the surface. In another STM experiment on epitaxially grown thin films, authors found superconducting properties that are different from the bulk behavior\cite{Lv16}---the critical temperature was somewhat larger and two gaps were detected in the tunneling conductance. Furthermore, a zero-bias peak appears in the center of the vortex cores, indicating the formation of vortex bound states \cite{Caroli64,Hess90,Guillamon08PRB}. Authors argue that these states could be topologically non-trivial, contrasting earlier results found in other superconducting materials. A recent report shows non-integer flux quantization in rings made of thin films of Bi$_{2}$Pd, suggesting the formation of a $\pi$ junction that authors associate with unconventional superconducting properties\cite{Li238}.

Our results provide new insight in this debate. The flux carried by single vortices can vary with respect to the flux quantum $\Phi_0$ in topological superconductors, due to enhanced stability of fractional vortices\cite{Vollhardt90}. The observation of vortices with a weaker magnetic field profile as the ones located on a defect in our case could be explained by their flux below $\Phi_0$, instead of a varying penetration depth. However, the change in $\lambda$ describes our results significantly better. Let us also remark that fractional flux quantization can also occur in long Josephson $0-\pi$ junctions, where the phase difference varies along the junction\cite{PhysRevB.70.174519,Kirtley1996}. Such junctions can form in superconductors with anisotropic gap structures or ferromagnetic inclusions. It is not clear how such a situation can be formed close to a linear defect in $\beta$-Bi$_{2}$Pd and lead to the observed increase in $\lambda$.

Thus, there is mounting evidence for the presence of triplet correlations in $\beta-$Bi$_2$Pd, particularly from experiments that are probing the surface. The question then arises of why we did not observe a half-integer flux quantum in our experiment. The answer might be that triplet correlations could form only close to the surface. Magnetic vortices, however, are threads through the bulk of the material and the surface properties might be masked by the flux quantization from the bulk. It would be very interesting to repeat our experiments in thin films or very thin layers of $\beta-$Bi$_2$Pd. The appearance of half-integer flux quanta is, in view of the recent reports \cite{Sakano15,Iwaya2017,Li238}, quite likely. Vortices might then carry a Majorana fermion, as proposed in Refs. \onlinecite{Caroli64,Hess90,Guillamon08PRB,Alicea2012,Beenakker2013}. In that case, the large intervortex distances found in the range of magnetic fields we study here should be very helpful to facilitate manipulation of vortices and braiding experiments. For example, a couple of vortices located in between lines could be easily moved around each other.

\color[rgb]{0,0,0}
\bibliographystyle{naturemag}

\end{document}